\documentclass[a4paper,11pt]{article}
\pdfoutput=1 % if your are submitting a pdflatex (i.e. if you have
\usepackage{jheppub} % for details on the use of the package, please
\usepackage{tikz}
\usetikzlibrary{decorations.pathmorphing}
\usetikzlibrary{shadows.blur}
\usetikzlibrary{intersections}
\usepackage[T1]{fontenc} % if needed
\usepackage{braket}
\def\Tr{\mathrm{Tr}}
\def\tr{\mathrm{tr}}

\def\be{\begin{equation}}
\def\ee{\end{equation}}
\def\ra{\rangle}
\def\la{\langle}
\usepackage{xcolor}\newcommand{\R}{\mathbb{R}}\renewcommand{\d}{\mathrm{d}}\renewcommand{\i}{i}
\newcommand{\BUket}[6]{%
 \begin{scope}[shift={(#1,#2)}, yscale={\ifnum#3=1 -#4\else #4\fi}, xscale=#4]
 \ifnum#6=1
 \draw[dotted, thick] (0,0) ellipse (.5 and .25);
 \draw[dotted, thick] (3,0) ellipse (.5 and .25);
 \draw[thick] (0.5,0) arc[start angle=0, end angle=180, radius=0.5];
 \draw[thick] (3.5,0) arc[start angle=0, end angle=180, radius=0.5];
 \else
 \draw[thick] (0,0) ellipse (.5 and .25);
 \draw[thick] (3,0) ellipse (.5 and .25);
 \fi

 \ifcase#5
 \draw[thick,dashed] (1.5,0) ellipse (.3 and .15); % 0 = dashed
 \or
 \draw[double,thick] (1.5,0) ellipse (.3 and .15); % 1 = thick double
 \or
 \draw[very thick,green] (1.5,0) ellipse (.3 and .15); % 2 = thick green
 \or
 \draw[thick] (1.5,0) ellipse (.3 and .15); % 3 = normal solid
 \fi

 \draw[thick] (3.5,0) arc[start angle=0, end angle=-180, radius=2];
 \draw[thick] (2.5,0) arc[start angle=0, end angle=-180, radius=1];
 \draw[thick, dashed] (1.2, 0) to[out = -90, in =80] (1,-1.5);
 \draw[thick, dashed] (1.8, 0) to[out = -90, in =100] (2,-1.5);
 \end{scope}%
}

\title{Wormholes and Averaging over N}

\author[]{Jonah Kudler-Flam and}
\author[]{Edward Witten}

\affiliation[]{School of Natural Sciences, Institute for Advanced Study, Princeton, NJ, USA}

\emailAdd{jkudlerflam@ias.edu}
\emailAdd{witten@ias.edu}

\abstract{The gravitational path integral produces an asymptotic expansion in $G_N$, a fact which is puzzling in the case of observables that are expected to fluctuate wildly. Wormholes  appear to compute ensemble averages of functions of such observables, though  in typical constructions of AdS/CFT, there are no parameters to average over except, in some examples, a single integer $N$. We introduce a procedure that we call ``Mellin averaging'' to define a sort of asymptotic average of a function of $N$.
We argue that Mellin averaging over $N$  may suffice to reproduce the apparent randomness seen in wormhole physics, provided that the dual theory admits an analytic continuation in $N$ and the relevant observables fluctuate on superpolynomially small scales in $N$. As a test case, we consider the spectral form factor in the regime where the double cone is believed to dominate the gravitational path integral and compare to a random matrix theory in which $N$ behaves as a continuous variable.  We also describe some toy models of analytic continuation in $N$: a qubit model that can be analytically continued in $N$, and an explicit construction of a deterministic function of $N$ that simulates a sequence of independent draws from a Gaussian ensemble.}

\begin{document} 
\maketitle
\flushbottom
\def\A{{\mathcal A}}
\def\var{{\mathrm {var}}}
\def\OO{{\mathbb{O}}}
\section{Introduction}
\label{sec:intro}

Wormholes are contributions to the gravitational path integral that involve a geometrical connection between widely separated regions of spacetime or between two otherwise disconnected spacetimes.
For quite some time, wormholes  have presented a puzzle for quantum gravity.  Arguments by Coleman~\cite{Coleman}, further developed
 by Banks, Klebanov and Susskind~\cite{BKS}, suggest that wormhole contributions introduce randomness, with an average over coupling constants.  This is in conflict with the expectation that a definite quantum gravity theory will give definite answers; for example,
there is no obvious source of randomness in string theory.  
  The puzzle has become more acute in recent years as interesting new applications of wormholes have emerged.  One type of example arises in the study of the spectral form factor, where a definite observable seems to behave as a family of Gaussian random variables~\cite{SSS}.   Another type of example emerged when advances in understanding the Page curve~\cite{AEMM,P} were interpreted in terms of wormholes~\cite{AHMST,PSSY}.    It appears that an average over many microscopic couplings is needed to produce the apparent randomness that comes from wormhole physics.

Perhaps the sharpest apparent conflict between wormhole physics and what we believe we understand from other points of view arises in the AdS/CFT correspondence.
This correspondence is believed to provide  a nonperturbative definition of quantum gravity in a world with a negative cosmological constant.   Depending on the asymptotic behavior of spacetime that is assumed, this nonperturbative definition is believed to depend on only a small number of parameters -- typically one or a few integers, and possibly a few continuously adjustable parameters.   Many important examples of AdS/CFT duality, such as the duality between M-theory in a world asymptotic to
 $\mathrm{AdS}_4 \times \mathrm{S}^7$ and a certain maximally supersymmetric conformal field theory in dimension 3~\cite{Aharony:2008ug}, depend only on a single integer $N$.   
 If wormhole physics reflects averaging over coupling constants, and in certain examples the only possible coupling is an integer $N$, we are forced to ask:
  is ``averaging over $N$'' sufficient to account for the apparent randomness associated to wormholes?\footnote{The proposal that averaging over $N$ may account for wormholes in AdS/CFT was put forward in~\cite{Schlenker:2022dyo} and recently proposed to be important for understanding closed universes in AdS/CFT~\cite{Liu:2025cml,Kudler-Flam:2025cki}. In \cite{Liu:2025ikq}, it was suggested that there may be an intrinsic way to separate the erratic and smooth pieces of $F(N)$ without explicit averaging, where the procedure was called a ``large-$N$ filter.'' Mellin averaging can be understood as an explicit proposal for such a filter.}

Naively,  the answer appears to be ``no,'' because  averaging over the values of a function that only depends on an integer $N$ does not have enough statistical power to account for wormhole physics.   After $N$ observations of a random variable, one will see fluctuations of
order $1/\sqrt N$, which would overwhelm -- and make it impossible to define --  the corrections  suppressed by powers of $1/N$ that can be computed from the gravitational path
integral. We will argue in this article that, nonetheless, averaging over $N$ may possibly account for the randomness that the gravitational path integral seems to generate via wormholes, but only if one makes certain fairly strong assumptions.    First, if $F(N)$ is an observable that in the gravitational path integral appears to behave as a random variable, then in our interpretation it is important that $F(N)$ is defined not just at integer values of $N$ but as a function of a continuous variable $N$ that is sufficiently well-behaved, in a sense we will make precise in section~\ref{mellin}.   A motivation for this assumption, apart from the fact that we will need it in order to make sense of wormhole physics, is that in the gravitational path integral, $N$ does appear to be a continuously variable parameter.\footnote{Chern-Simons theory in three dimensions provides an example of analytic continuation away from integers.   This theory depends on an integer parameter $k$, but the expectations of Wilson
 loops in $\R^3$ can naturally be defined as holomorphic functions of $k$.   Analytic continuation in $k$  was originally understood as a statement about the Jones polynomial of a knot and the Knizhnik-Zamolodchikov equation~\cite{Kohno,TK}. However, the analytic continuation in $k$ has a natural explanation from the 
 Chern-Simons path integral~\cite{Witten}.}
 Second, to account for what is seen in wormhole physics, it must be that if $F(N)$ is an observable that has important wormhole contributions, then, as a function of a real variable $N$, $F(N)$ is wildly oscillatory with oscillations on a scale that vanishes for large $N$  faster than any power of $1/N$.  Roughly speaking, the importance of this is that averaging over nonperturbatively fast oscillations has much more statistical power than simply averaging over the values of $F(N)$ for integer $N$.  For a certain class of functions that satisfy these conditions, we will propose a procedure of ``Mellin averaging'' that uses a Mellin transform  to interpret $F(N)$  -- at integer values of $N$, where it is originally defined -- as a sequence of random variables. While we refer to this procedure as Mellin ``averaging,'' it is not  averaging in a conventional sense. It is a procedure to define a sort of asymptotic average. Mellin averaging does not produce a function of $N$ but only a $1/N$ expansion near $N=\infty$.   We believe that that is consistent with what is needed, since such an expansion is what the gravitational path integral provides.   It is not clear that there is a natural notion of ``averaging over $N$'' that converts a wildly oscillatory function of $N$ into a smooth one, but the gravitational path integral as usually understood produces only an asymptotic expansion.

As a prototype for an AdS/CFT observable that is strongly affected by wormholes, we will consider an analytically continued version of the CFT partition function, namely $Z_N(\beta+\i T)=\Tr\,e^{-(\beta+\i T)H_N}$,  where $H_N$ is the CFT Hamiltonian, which for simplicity we assume depends on only a single integer $N$,
and where $T\in\R$, $\beta>0$.   The reason to focus on this example is that the relevant wormhole, namely the  two-boundary ``double cone,''  is known and relatively well understood~\cite{SSS,CIM}.   The double cone is a connected (complexified) spacetime whose boundary has two disconnected components
with opposite orientation, so in the gravitational path integral it seems to make a ``connected'' contribution to $|Z_N(\beta+\i T)|^2$.   The function $|Z_N(\beta+\i T)|^2$ is
known as the spectral form factor.\footnote{In chaos theory, the spectral form factor is often defined with $\beta=0$, but in AdS/CFT, the behavior of the spectrum at
high energies is such that $\beta>0$ is needed as a regulator.}    For sufficiently large $T$, $|Z_N(\beta+\i T)|^2$, computed on the double cone 
spacetime,\footnote{Actually, technically one has to replace  $Z_N(\beta+\i T)$ with a microcanonical analog.  We will incorporate this in a more detailed discussion 
in section~\ref{sec:SFFADSCFT}.} is much bigger
than the ``disconnected'' contribution that one gets by computing separately the two factors of $|Z_N(\beta+\i T)|^2=Z_N(\beta+\i T) \overline{ Z_N(\beta+\i T)}$,
each on its own connected spacetime (and therefore, overall, on a disconnected spacetime).  
This does not make much sense if one assumes that $Z_N(\beta+\i T)$, as computed in the gravitational path integral, is a definite complex number.
It  has been interpreted to mean
that the gravitational path integral is computing some sort of average, and that   the average of $|Z_N(\beta+\i T)|^2$, for large $T$,  is much greater
than the absolute value squared of the average of $Z_N(\beta+\i T)$:
\be\label{compav}\la|Z_N(\beta+\i T)|^2\ra\gg |\la Z_N(\beta+\i T)\ra|^2. \ee
A motivation for this idea is the observation~\cite{SMany} that if gravity can be understood as a chaotic quantum system~\cite{SS}, then $Z_N(\beta+\i T)$, for large $T$,
should depend erratically on $T$, with wild fluctuations on a scale of order $\beta$. 
The  gravitational path integral -- at least in the semiclassical sense in which we study it in practice -- does not exhibit such wild fluctuations,
 suggesting that it actually computes some sort of average with the fluctuations washed out.

One can similarly consider more general averages $\la Z_N(\beta+\i T)^p Z_N(\beta-\i T)^q\ra$ for any positive integers $p,q$.   Assuming that for sufficiently
large $T$, the dominant 
contributions come from the disjoint union of multiple double cones (and therefore that these averages  are negligible unless $p=q$), and not from additional
wormholes with more than two boundaries,
it has been argued~\cite{SSS} that at late times, $Z_N(\beta+\i T) $ behaves in the gravitational path integral as  a 
Gaussian random variable.     One may further surmise that $Z_N(\beta+\i T)$ and $Z_{N'}(\beta+\i T)$, for $N\not=N'$, are largely
independent of each other as random variables, because they are 
 computed in different chaotic quantum systems.  This is supported by computations  and will be discussed in 
 section~\ref{sec:SFFADSCFT}.

But what sort of averaging is going on here?   One observation is that, while $Z_N(\beta+\i T)$ is expected to be 
wildly oscillatory for large $T$, averaging it over a 
range of $T$ large compared to $\beta$ is expected (by analogy with general results about quantum chaos) to give a 
smooth function of $T$ that  behaves approximately as a Gaussian random variable.
However, as an interpretation of the double cone contribution, this observation has two limitations.
First, in AdS/CFT, we are entitled to study $Z_N(\beta+\i T)$ for fixed $\beta$ and $T$, and we still seem to get a wormhole contribution leading to the inequality
\eqref{compav} and Gaussian behavior assuming that multi-boundary wormholes are not important.  So an explanation that relies on averaging over $T$ does
not fully account for the phenomenon that has to be explained.  Second, although it is true that averaging over a reasonable
range of $T$ does give a variable that behaves in an approximately Gaussian fashion, this actually does not produce a variable that is Gaussian with the precision that seems to come from the gravitational path integral. 
In the gravitational path integral, when $T$ is sufficiently large compared to $N$,
corrections to Gaussian behavior are believed to be exponentially small (in $N$),
coming from connected wormhole geometries with more than two asymptotic boundaries.    In a framework of averaging over $T$, to
 make the corrections to Gaussian behavior
exponentially small, one would have to average over a range of $T$ that is exponentially large compared to $\beta$. This is because the oscillations in time are only on a scale of $O(\beta)$. This amount of time averaging would be needed to have enough statistical power to make the $1/N$ corrections well-defined.
That is not usually considered and is not really desirable, because the ``averaged'' behavior of the spectral form factor
is not constant over such a long range of $T$; famously, the spectral form factor is linearly growing in $T$~\cite{SMany}. 

As an alternative, we will in this article treat $\beta$ and $T$ as fixed real numbers and explore the hypothesis that some version of ``averaging over $N$''
is sufficient to interpret $Z_N(\beta+\i T)$ as a  random variable that becomes Gaussian for large $T$.    As already noted, the version of averaging over 
$N$ that we consider assumes that $Z_N(\beta+\i T)$, for fixed values of $\beta$ and $T$,  can be treated as an analytic function of $N$ with some favorable properties.   And the output of the averaging is only an asymptotic expansion in powers of $1/N$ near $N=\infty$, not a smooth function of $N$.

The contents of this article are as follows.   In section~\ref{mellin} we explain our proposal for averaging based on the Mellin transform.
In section~\ref{sec:SFFADSCFT}, we describe the double cone and the ``spectral form factor'' $|Z_N(\beta+\i T)|^2$ in more detail.  We also discuss correlations
between the partition functions $Z_N(\beta+\i T)$ for different values of $N$ or $T$, and show that for large $N$ or large $T$, these correlations become extremely small. 
And we discuss the question of how large $T$ must be so that the double cone dominates.   This question turns out to be critical for the viability of our
interpretation of the averaging produced by the double cone.
 In section~\ref{sec:loop} (with further details in some appendices), we use random
matrix theory to explore what is a reasonable hypothesis for the behavior of $Z_N(\beta+\i T)$ as a function of $N$ and $T$.   Random matrix theory is useful
here not just because it gives a good guide to many properties of chaotic quantum systems, but because it provides a framework in which $N$ can be treated as a 
continuous variable.   In section~\ref{toymodels}, as a reality check, we construct toy models of deterministic functions $F(N)$ whose values for integer $N$
behave, under Mellin averaging, as Gaussian random variables.  And in section~\ref{toyanalytic},
we explain how to analytically continue some toy models of a black hole as a function of $N$.

\section{Averaging Over N and the Mellin Transform}\label{mellin}

Consider an example of AdS/CFT duality that depends only on an integer $N$ and no other parameters.  
We want to ask whether an observable $F(N)$ in this theory can be ``averaged over $N$'' to account for what is seen in the gravitational path integral via wormhole 
effects.
A basic problem, already mentioned in the introduction, is that an integer variable $N$ does not have enough values.   In more detail, suppose that
an observable $F(N)$ in AdS/CFT duality behaves erratically as a function of $N$ and that the gravitational path integral is really producing
an averaged value $\la F(N)\ra$ of this variable.
The output of the gravitational path integral has an expansion in powers of Newton's constant $G$.  In CFT language, this is a $1/N$ expansion:\footnote{For simplicity, we assume
that the relation between $G$ and $N$ is that $G\sim 1/N$, though in actual examples of AdS/CFT duality, in general $G\sim 1/N^\lambda$, with a model-dependent
exponent $\lambda$.   Also for simplicity, we assume that $\la F(N)\ra$  has a simple $1/N$ expansion, though in general one might have something
like $\la F(N)\ra \sim \exp(-N I_0)N^{-\gamma}\left(f_0+\frac{f_1}{N}+\cdots\right)$, where $I=N I_0$ is the action of a dominant classical solution, and the factor 
$N^{-\gamma}$ comes from a one-loop determinant. In these cases, we can strip off the simple prefactor to apply averaging to the simple $1/N$ expansion.} 
\be\label{expn} \la F(N)\ra =f_0+\frac{f_1}{N}+\frac{f_2}{N^2} +\cdots.\ee
  In the presence of wormholes, the gravitational path integral likewise produces an expansion of $F(N)^2$
and of higher powers of $F(N)$.   For example,
\be\label{secondpower}\la F(N)^2\ra =g_0+\frac{g_1}{N}+\frac{g_2}{N^2}+\cdots.\ee
If $F(N)$ is, for example, the real or imaginary part of $Z_N(\beta+\i T)$, for $T$ large enough that the disconnected contribution to the spectral form factor
is negligible, then  $\la F(N)\ra $ is identically zero, but the double cone wormhole gives a non-trivial $1/N$ expansion of $\la F(N)^2\ra$.    This is
interpreted to mean that $F(N)$ oscillates erratically as a function of $N$, 
and the gravitational path integral averages out these oscillations to give zero.   But of course,
if $F(N)$ oscillates erratically as a function of $N$, then $F(N)^2$ generically also oscillates erratically (around a nonzero mean determined by the variance of $F(N)$).   The
gravitational path integral is somehow averaging over these oscillations to produce an asymptotic expansion of $\la F(N)^2\ra$  in powers of $1/N$.   We want to
understand how this might work.

There is no mystery about how gravity might be defining the leading coefficient:
\be\label{leading} g_0=\lim_{N\to\infty}\frac{1}{N}\sum_{N'=1}^N F(N')^2.\ee   This limit will exist if the observables $F(N)$ are drawn independently from a family of
distributions that has a large $N$ limit.   However, if $F(N)$ behaves as a random variable, then $F(N)^2$ also 
behaves as a random variable with a non-trivial variance.
So  $g_0(N)=\frac{1}{N}\sum_{N'=1}^N F(N')^2$ will fluctuate about its limiting value with a typical amplitude of order $1/\sqrt N$.
This exceeds in magnitude the subleading terms $g_1/N+g_2/N^2+\cdots$ in the $1/N$ expansion of $\la F(N)^2\ra $, so with no information except the sequence
of values $F(1), F(2),\dots$,  it will not be possible to define the subleading
coefficients $g_1,g_2,\dots$.   

A possible way out is to assume that the observables $F(N)$ are really the values at integers of a function $F(N)$ of a real variable. Actually, the averaging
procedure that we will define is most natural if 
any function $F(N)$ that is significantly affected by wormholes
fluctuates wildly on a scale that, for large $N$, is smaller than any power of $1/N$.   Imagine more specifically that for some function $\delta(N)$
that vanishes for large $N$ faster than any power of $1/N$, the quantities $F(N)$ and $F(N')$ behave as independent random variables if $|N-N'|>\delta(N)$.  If
that is so, then the number of independent values of $F(N)$ for $N$ less than some large number $N_{\mathrm{max}}$  grows faster than any power of $N_{\mathrm{max}}$.  Then fluctuations in averages of functions such as $F(N)$, $F(N)^2$, etc., will vanish faster than any power of $1/N$, and potentially will not obstruct
the existence of an averaging procedure that defines averaged $1/N$ expansions of these functions.   In the definition of averaging that we will give in this
section, we do not need to assume that $F(N)$ has the detailed behavior just described.  But heuristically, it must have such behavior if the averaging
procedure is to reproduce what is found in wormhole physics. In section~\ref{toymodels}, we will see that this expectation is satisfied in toy models.

What can the averaging procedure be?
The most obvious idea is to literally average over $N$ with some window function 
$k(N,N')$ and define
\be\label{zincom}\la F(N)\ra=\int \d N'\,k(N,N') F(N'). \ee   
If $k(N,N')$ is slowly varying, such averaging will eliminate rapid fluctuations of $F(N)$ and produce a smoother average function 
$\la F(N)\ra$.   However, there are some basic problems with this notion
of averaging.
One problem  is that no canonical choice of the window function $k(N,N')$ is apparent.
Even more serious  is the following.  There are plenty of observables in AdS/CFT that do {\it not} behave in the gravitational path integral as fluctuating
random variables; these observables have straightforward $1/N$ expansions that are not affected by wormholes.
If $F(N)$ has this property, then there is no need for any non-trivial averaging, and we want any averaging procedure to act trivially.   That will not be the case if
we average in the straightforward sense of \eqref{zincom}.   Applied to a function $F(N)=f_0+f_1/N+\cdots$ that has a $1/N$ expansion, the averaging
procedure of \eqref{zincom} will generically give $\la F(N)\ra \not=F(N)$.

We will instead consider an averaging procedure based on the Mellin transform.   Consider a
function $F(N)$ that is bounded by a power of $N$ for large $N$.  Its Mellin transform is defined by\footnote{If confusion is unlikely, we write just $M(s)$ rather than $M_F(s)$.}
\be\label{elmet} M_F(s)=\int_1^\infty \d N\, N^{s-1} F(N). \ee
The  integral converges if ${\rm Re}\,s$ is sufficiently negative. 
The procedure that we will follow  with the Mellin transform has the property
that it defines $\la F(N)\ra$ only as an asymptotic expansion in powers of $1/N$, not as an actual function of $N$.  The class of functions to which our
averaging procedure applies are functions $F(N)$ with the property that, after analytic continuation from the region of the $s$-plane where the integral
\eqref{elmet} converges,  $M_F(s)$ is holomorphic on the real axis except for simple poles at non-negative integers.
If that is so, we define
\be\label{zelme} f_k=-{\mathrm{Res}}_{s=k} \, M_F(s) \ee
and
\be\label{welme} \la F(N)\ra=\sum_{k=0}^\infty  \frac{f_k}{N^k}.\ee
We call this the Mellin average of $F(N)$.
The motivation for the definition is that the Mellin transform of the function $W_r(N)=N^{-r}$ is holomorphic except for a simple pole at $s=r$ with residue $-1$:
\be\label{pelme} M_{W_r}(s)=\int_1^\infty \d N N^{s-1-r}=\frac{1}{r-s}.\ee
Hence, the Mellin average of $1/N^r$ is precisely $1/N^r$, and more generally any function with a $1/N$ expansion
is unchanged by the averaging:   if $F(N)=f_0+f_1/N+\cdots$ then
\be\label{welcox}\la F(N)\ra=F(N). \ee

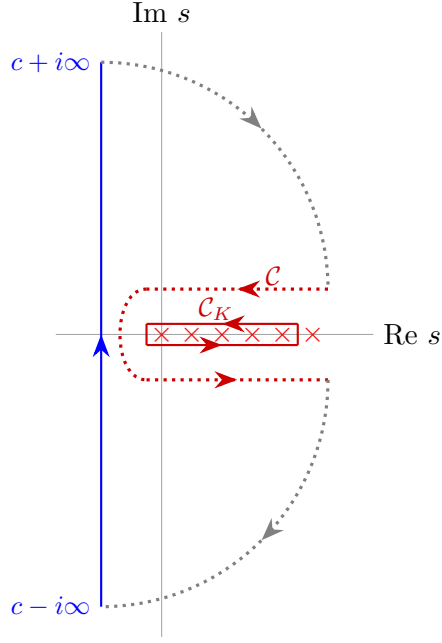
\begin{figure}
    \centering
      \usetikzlibrary{arrows.meta}
  \usetikzlibrary{decorations.markings}

\begin{tikzpicture}[
    scale=0.4,
    >={Stealth[length=3mm]},
    thick,
    midarrow/.style={
      postaction={decorate},
      decoration={markings,
        mark=at position 0.5 with {\arrow{Stealth[length=3mm]}}}
    }
  ]

  % ===== Geometry =====
  \def\cval{-2}
  \def\xL{-0.5}
  % --- Inner contour C_K (solid) ---
  \def\xR{4.5}
  \def\yT{0.35}
  \def\yB{-0.35}
  % --- Outer semi-infinite contour C (dotted) ---
  \def\Xfar{5.5}    % must be > K to be distinct from C_K
  \def\yTc{1.5}
  \def\yBc{-1.5}
  % --- Quarter circles attach to C ---
  \def\Rad{7.5}     % = Xfar - cval
  \def\Htop{9}      % = yTc + Rad
  \def\Hbot{-9}     % = yBc - Rad

  % ===== Axes =====
  \draw[thin, gray!70] (-3.5, 0) -- (7, 0)
        node[right, black] {$\operatorname{Re}\,s$};
  \draw[thin, gray!70] (0, -10) -- (0, 10)
        node[above, black] {$\operatorname{Im}\,s$};

  % ===== Poles at 0, 1, ..., K =====
  \foreach \n in {0,1,2,3,4,5} {
    \node[red] at (\n,0) {$\times$};
  }
  % \node[below=8pt, font=\small] at (1,0) {$1$};
  % \node[below=8pt, font=\small] at (2,0) {$2$};
  % \node[below=8pt, font=\small] at (3,0) {$3$};
  % \node[below=8pt, font=\small] at (4,0) {$K$};
  % \node[font=\tiny] at (5.0,0) {$\cdots$};

  % ===== Bromwich contour (going up) =====
  \draw[blue, midarrow] (\cval, \Hbot) -- (\cval, \Htop);
  \node[blue, font=\small, left] at (\cval, \Hbot) {$c-i\infty$};
  \node[blue, font=\small, left] at (\cval, \Htop) {$c+i\infty$};

  % ===== Inner solid rectangular contour C_K =====
  \draw[red!80!black, midarrow]  (\xL, \yT)--(\xR, \yT) ;
  \draw[red!80!black]           (\xL, \yT) -- (\xL, \yB);
  \draw[red!80!black, midarrow] (\xR, \yB)-- (\xL, \yB) ;
  \draw[red!80!black]           (\xR, \yB) -- (\xR, \yT);
  \node[red!80!black, font=\small] at (1.8, 0.85) {$\mathcal{C}_K$};

  % ===== Outer dotted semi-infinite contour C =====
  \draw[red!80!black, dotted, very thick, midarrow]   (\xL, \yTc)--(\Xfar, \yTc);
  \draw[red!80!black, dotted, very thick]           (\xL, \yTc) to[out = 180, in = 180](\xL, \yBc);
  \draw[red!80!black, dotted, very thick, midarrow] (\Xfar, \yBc)--(\xL, \yBc) ;
  \node[red!80!black, font=\small] at (3.7, 1.95) {$\mathcal{C}$};

  % ===== Quarter-circle arcs at infinity (dotted gray) =====
  % Top quarter: center (cval, yTc), from angle 90° to 0° (CW)
  \draw[gray, dotted, very thick, midarrow]
    (\cval, \Htop) arc[start angle=90, end angle=0, radius=\Rad];
  % Bottom quarter: center (cval, yBc), from angle 0° to -90° (CW)
  \draw[gray, dotted, very thick, midarrow]
    (\Xfar, \yBc) arc[start angle=0, end angle=-90, radius=\Rad];

\end{tikzpicture}
    \caption{The red $\times$'s represent poles of $M_F(s)$. The blue contour is the one used for the inverse Mellin transformation while the dotted red contour, $\mathcal{C}$, is the one used for Mellin averaging. The integral over the gray contour equals the difference. The solid red contour, $\mathcal{C}_K$, is the convergent contour used to define the asymptotic series truncated to order $K$.}
    \label{fig:contour}
\end{figure}

Mellin averaging can be understood as a contour deformation in the $s$-plane. The Mellin transform can be inverted for suitable conditions on $M_F(s)$ via the contour integral
\begin{align}
    F(N) = \frac{1}{2\pi i}\int_{c-i\infty}^{c+i\infty} ds N^{-s} M_F(s), \quad N > 1,
\end{align}
where $c<0$ and the lower bound on $N$ arises from the lower limit of the integral in the Mellin transform. Mellin averaging is equivalent to deforming this contour to wrap the positive real axis as in $\mathcal{C}$ of figure \ref{fig:contour}
\begin{align}
    \langle F(N)\rangle = \frac{1}{2\pi i}\int_{\mathcal{C}} ds N^{-s} M_F(s).
\end{align}
Assuming analyticity of $M_F(s)$ off the positive real axis, $\langle F(N)\rangle$ differs from $F(N)$ only by the contribution from the arcs at infinity (the gray dotted lines of the figure). For $F(N)$ that already had good asymptotic expansions, this contribution is trivial, but for $F(N)$ that behave erratically at large $N$, these contributions are significant. One conceptual benefit of expressing $\braket{F(N)}$ as a contour integral, rather than the formal sum of \eqref{welme}, is that the powers of $N$ appearing in series do not need to be known a priori. For example, an overall factor of $N^{-\gamma}$ coming from a one-loop determinant would simply shift the locations of the poles by $\gamma$.

Because the residue series at $s=0,1,2,\dots$ is generically only asymptotic, the integral over the full semi-infinite contour $\mathcal{C}$ does not converge in general. We therefore define the order-$K$ truncation $\braket{F(N)}_K$ as the integral over the closed contour $\mathcal{C}_K$ that wraps only the poles at $s=0,1,2,\dots, K$.

We will describe a few simple examples of Mellin averaging, and defer a more detailed discussion of examples that better simulate the behavior of gravity to
section~\ref{toymodels}.   As a first example, consider the function $F(N)=\cos(aN)$, with a nonzero real constant $a$.   Intuitively, the average of this function should
vanish, and that is the answer that comes from Mellin averaging, because the Mellin transform
\be  M_F(s)=\int_1^\infty \d N\, N^{s-1} \cos(a N) \ee
is an entire function\footnote{This integral is actually a hypergeometric function, but instead of invoking that fact, we will explain a general strategy to show
that such an integral defines an entire function of $s$.} of $s$.    To prove that, we use the identity $\cos(aN)=(-1)^r a^{-2r}\frac{\d^{2r}}{\d N^{2r}}\cos (aN)$, with arbitrary positive integer $r$.
Inserting this identity in the definition of the Mellin transform, we get
\be\label{elme} M_F(s)=(-1)^r a^{-2r}\int_1^\infty\d N\,N^{s-1}  \frac{\d^{2r}}{\d N^{2r}}\cos (aN).\ee
Now we integrate by parts $2r$ times.   We pick up many surface terms at $N=1$, but they are all entire functions of $s$ and do not contribute to the poles
of the Mellin transform.   For sufficiently negative ${\rm {Re}}\, s$, there is no surface term at $N=\infty$.   Apart from the surface terms, we are left with a multiple
of the integral $\int_1^\infty \d N\, N^{s-1-2r}\cos(aN)$.   For any $s$, we can pick $r$ so that this integral converges.   So $M_F(s)$ is an entire function of $s$,
and therefore in the sense of Mellin averaging, $\la F(N)\ra=0$.

We can also consider $\la F(N)^2\ra$.   We have $\cos^2(aN)=\tfrac{1}{2}+\tfrac{1}{2}\cos (2 aN)$.  The Mellin average of $\cos(2aN)$ vanishes, by the
reasoning just explained.   And, as a special case of the statement that Mellin averaging acts trivially on functions with a $1/N$ expansion, the Mellin
average of $\tfrac{1}{2}$ is $\tfrac{1}{2}$. So the Mellin average of $\cos^2(aN)$ is $\tfrac{1}{2}$.  This is in accord with the fact that the average of $\cos^2(aN)$, computed by treating $aN$ as a random angle,
is $\tfrac{1}{2}$.

Another instructive example is $F(N)=\left(1+\frac{1}{N}\right) \cos(aN)$.   We will leave it to the reader to show that for this function, Mellin averaging gives
$\la F(N)\ra = 0$ and $\la F(N)^2\ra=\tfrac{1}{2}\left(1+\frac{1}{N}\right)^2$.   This example is instructive because it illustrates that, for an oscillatory function 
$F(N)$ whose Mellin average vanishes,  Mellin averaging succeeds in 
extracting higher terms in the $1/N$ expansion of $\la F(N)^2\ra$.

In section~\ref{toymodels}, we will discuss some more elaborate examples.   The goal will be to construct a toy function $F(N)$ whose values for integer $N$
behave as independent draws from a Gaussian distribution.

Returning to classical averaging, the average $g_0$ defined in \eqref{leading} is obviously unchanged if we change the first $N_0$ values of $F(N)$ for any $N_0$; it only depends on the asymptotic behavior of the series.   The analog of this for Mellin averaging is that the poles of the Mellin transform $M_F(s)$ are unchanged if we define the Mellin transform by integration over the region $N\geq N_0>1$ rather than $N\geq 1$:
\be\label{truncated} M_F^{(N_0)}(s)=\int_{N_0}^\infty \d N \,N^{s-1} F(N).\ee  Changing the lower limit of the integral only changes the Mellin transform by an entire function of $s$.

In this paper, we only aim to address the perturbative expansion around a single saddle of the gravitational path integral using Mellin averaging. However, the gravitational path integral is generally a sum over saddle points with different classical actions and a perturbative series around each saddle point. The possibility of reproducing subleading saddles from the CFT remains a puzzle that we do not claim to solve. In the following, we briefly comment on an extension of the Mellin transform that can map erratic functions of $N$ to formal transseries of the form\footnote{For simplicity, we assume no additional powers of $N$ from one-loop and no logarithms, though the generalizations are straightforward.}
\begin{align}
    \sum_\alpha  e^{-N I_\alpha}\sum_{k = 0}^{\infty}\frac{f_k^{(\alpha)}}{N^k}.
\end{align}
In each $\alpha$ sector, the perturbative series is only asymptotic. We cannot perform some ``optimal truncation'' on them because these introduce non-perturbative errors that end up mixing sectors. We also do not want to assume that there is any resurgent structure in this transseries. In particular, we do not assume that the perturbative series around the leading saddle determines the locations $I_\alpha$ or coefficients $f_k^{(\alpha)}$ of subleading saddles. We thus can only consider the output of the gravitational path integral as a formal sum of asymptotic series, with each sector's data $\{I_\alpha,  f_k^{(\alpha)}\}$ treated as independent input. On the boundary, there is a genuine function of $N$, $F(N)$, because there is nothing wrong with computing an observable at fixed $N$. 

We can apply a Mellin--Laplace transform to $F$ 
\begin{align}
    H_F(s,p)\equiv \int_1^\infty dN\, N^{s-1}\, e^{-pN}\, F(N),
\end{align}
which produces a genuine function of $(s,p)$ whose analytic structure we will use to extract the transseries data. To see what analytic structure to look for, consider the termwise Mellin--Laplace image of the formal transseries $\la F(N)\ra$, which takes the form
\begin{align}
    \sum_\alpha \sum_{k \ge 0} f_k^{(\alpha)}\, \Gamma(s - k)\, (p + I_\alpha)^{k - s} \;+\; (\text{analytic in } p).
\end{align}
The extraction proceeds in two steps. First, the singularities of $H_F(s,p)$ in the $p$-plane at fixed generic $s$ are located at $p = -I_\alpha$, one for each sector $\alpha$, so the classical actions are read off as minus the branch-point positions,
\begin{align}
    \{I_\alpha\} = \{-p_* : p_* \text{ is a singularity of } H_F(s,\cdot)\}.
\end{align}
Second, expand $H_F(s,p)$ locally at $p = -I_\alpha$ as
\begin{align}
    H_F(s,p) = \sum_{k \ge 0} A_k^{(\alpha)}(s)\, (p + I_\alpha)^{k - s} \;+\; R_\alpha(s,p),
\end{align}
where $R_\alpha(s,p)$ is analytic in $p$ at $p = -I_\alpha$. Matching against the termwise transform gives $A_k^{(\alpha)}(s) = f_k^{(\alpha)}\, \Gamma(s - k)$, hence
\begin{align}
    {\; f_k^{(\alpha)} = \operatorname*{Res}_{s = k} A_k^{(\alpha)}(s) \;}.
\end{align}
Combining these, we construct the formal transseries
\begin{align}
    \la F(N)\ra = \sum_\alpha e^{-N I_\alpha} \sum_{k = 0}^{\infty} \frac{f_k^{(\alpha)}}{N^k}.
\end{align}
Note that this procedure does not average out phases linear in $N$. For example, if $F(N)=\cos(aN)$ it finds singularities at $p=\pm \i a$ and returns $F(N)$ itself, unlike the Mellin average of section~\ref{mellin}, which vanishes. One may expect that this procedure can only match the gravitational path integral if $F(N)$ is oscillating on scales that are even smaller than exponential so that there is enough statistical power to resolve the exponentially suppressed terms. Concrete analysis of this proposal is outside the scope of this work.

\section{The Spectral Form Factor in AdS/CFT}
\label{sec:SFFADSCFT}
\begin{figure}
    \centering
\begin{tikzpicture}[scale=.7]

% Axes
\draw[->, very thick] (0,0) -- (13.5,0) node[right, font=\LARGE] {$T$};
\draw[->, very thick] (0,0) -- (0,9.0) node[above, font=\LARGE] {$|Z_{\Delta,N}|^2$};

% Tick marks on x-axis
\draw[very thick] (5.426, 0.10) -- (5.426, -0.10)
  node[below, font=\LARGE] {$T_{\mathrm{dip}}$};
\draw[very thick] (10.439, 0.10) -- (10.439, -0.10)
  node[below, font=\LARGE] {$T_{H}$};

% Tick marks on y-axis
\draw[very thick] (0.10, 8.000) -- (-0.10, 8.000)
  node[left, font=\LARGE] {$e^{2S}$};
\draw[very thick] (0.10, 3.362) -- (-0.10, 3.362)
  node[left, font=\LARGE] {$e^{S}$};

% Dashed reference lines
\draw[dashed, gray, thick] (5.426, 0) -- (5.426, 1.121);
\draw[dashed, gray, thick] (10.439,  0) -- (10.439,  3.362);
\draw[dashed, gray, thick] (0, 8.000) -- (0.8, 8.000);
\draw[dashed, gray, thick] (0, 3.362) -- (10.439, 3.362);

% Erratic single realization (ramp + plateau, from GUE data)
\draw[gray!80, thin] (5.426,0.548)
  -- (5.491,1.774)
  -- (5.556,0.000)
  -- (5.621,3.518)
  -- (5.686,1.696)
  -- (5.751,0.312)
  -- (5.816,0.000)
  -- (5.881,2.149)
  -- (5.947,1.286)
  -- (6.012,0.858)
  -- (6.077,0.000)
  -- (6.142,1.351)
  -- (6.207,0.000)
  -- (6.272,2.193)
  -- (6.337,1.198)
  -- (6.402,0.000)
  -- (6.467,0.000)
  -- (6.533,0.000)
  -- (6.598,2.154)
  -- (6.663,3.113)
  -- (6.728,0.444)
  -- (6.793,2.145)
  -- (6.858,0.000)
  -- (6.923,0.000)
  -- (6.988,2.269)
  -- (7.053,3.056)
  -- (7.119,0.840)
  -- (7.184,1.114)
  -- (7.249,1.793)
  -- (7.314,2.485)
  -- (7.379,2.984)
  -- (7.444,2.363)
  -- (7.509,5.200)
  -- (7.574,0.883)
  -- (7.639,0.000)
  -- (7.705,3.225)
  -- (7.770,2.359)
  -- (7.835,3.883)
  -- (7.900,2.488)
  -- (7.965,4.030)
  -- (8.030,3.724)
  -- (8.095,3.557)
  -- (8.160,1.434)
  -- (8.225,1.003)
  -- (8.290,3.087)
  -- (8.356,2.113)
  -- (8.421,2.454)
  -- (8.486,2.712)
  -- (8.551,2.003)
  -- (8.616,0.608)
  -- (8.681,3.950)
  -- (8.746,2.129)
  -- (8.811,3.514)
  -- (8.876,0.000)
  -- (8.942,3.052)
  -- (9.007,1.673)
  -- (9.072,2.997)
  -- (9.137,1.744)
  -- (9.202,1.081)
  -- (9.267,3.393)
  -- (9.332,2.700)
  -- (9.397,4.284)
  -- (9.462,2.798)
  -- (9.528,3.670)
  -- (9.593,0.000)
  -- (9.658,1.649)
  -- (9.723,4.036)
  -- (9.788,1.477)
  -- (9.853,2.823)
  -- (9.918,1.306)
  -- (9.983,4.414)
  -- (10.048,4.320)
  -- (10.114,1.017)
  -- (10.179,3.245)
  -- (10.244,3.523)
  -- (10.309,4.439)
  -- (10.374,2.765)
  -- (10.439,2.820)
  -- (10.504,3.097)
  -- (10.569,0.000)
  -- (10.634,2.492)
  -- (10.699,3.035)
  -- (10.765,1.641)
  -- (10.830,0.000)
  -- (10.895,3.177)
  -- (10.960,3.424)
  -- (11.025,2.971)
  -- (11.090,2.271)
  -- (11.155,1.423)
  -- (11.220,3.767)
  -- (11.285,2.540)
  -- (11.351,1.076)
  -- (11.416,3.292)
  -- (11.481,1.612)
  -- (11.546,2.360)
  -- (11.611,3.105)
  -- (11.676,3.742)
  -- (11.741,0.000)
  -- (11.806,0.644)
  -- (11.871,4.080)
  -- (11.937,0.000)
  -- (12.002,2.343)
  -- (12.067,3.326)
  -- (12.132,5.122)
  -- (12.197,4.369)
  -- (12.262,0.610)
  -- (12.327,4.020)
  -- (12.392,4.223)
  -- (12.457,1.665)
  -- (12.523,2.018)
  -- (12.588,3.671)
  -- (12.653,4.439)
  -- (12.718,3.416)
  -- (12.783,3.189)
  -- (12.848,3.025)
  -- (12.913,1.224)
  -- (12.978,1.962);

% Smooth average (ramp + plateau, from GUE data)
\draw[blue, line width=2.5pt, line cap=round] (5.426,1.121)
  -- (10.439,3.362)
  -- (13.2,3.362);

% Cartoon smooth dip (bezier), drawn on top
\draw[gray!80, thin]
  (0, 8.000)
  .. controls (1.356, 8.000) and (4.069, 1.921) ..
  (5.426, 1.121);
\draw[blue, line width=2.5pt, line cap=round]
  (0, 8.000)
  .. controls (1.356, 8.000) and (4.069, 1.921) ..
  (5.426, 1.121);

\end{tikzpicture}
\caption{A cartoon of the microcanonical spectral form factor in a chaotic theory. Before $T_{\mathrm{dip}}$, $|Z_{\Delta,N}|^2$ smoothly decays from its initial maximum, after which it oscillates erratically in time during the ``ramp,'' where it is, on average, growing linearly with $T$. At the Heisenberg time, $T_H$, the linear growth saturates to a ``plateau'' that is also highly erratic but constant on average.}
    \label{fig:SFF}
\end{figure}
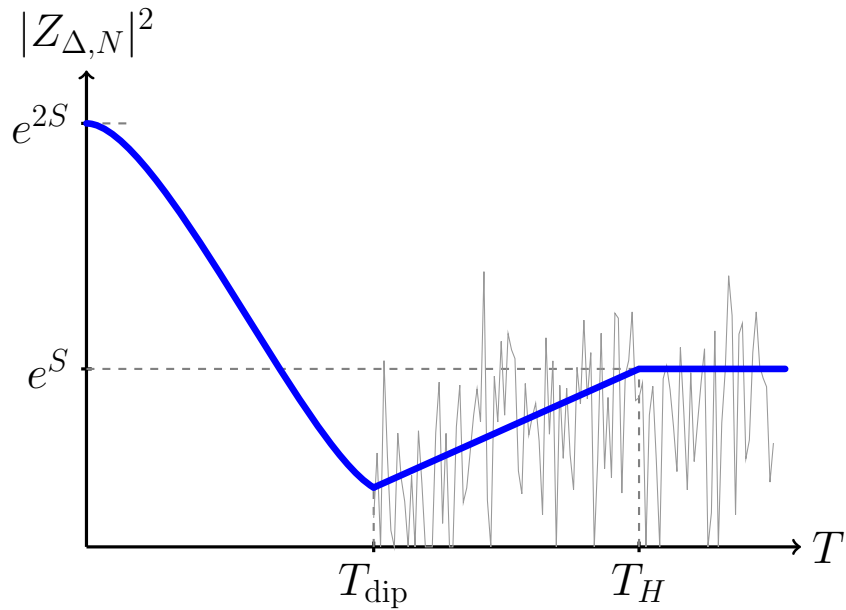

As explained in the introduction, the complex time partition function is a prototype of an observable whose analysis involves wormholes.  Actually, as first explained in \cite{SSS}, it is useful to 
consider a microcanonical observable in which the trace that defines the partition function is restricted to an energy band.
Specifically we define
\begin{align}
     Z_{\Delta,N}(T)\equiv \Tr[w_{\Delta}(H_N)e^{-\i H_NT}]
\end{align}
where $w_{\Delta}(H)$ is a smooth microcanonical window function that we take to be Gaussian
\begin{align}
    w_\Delta(x) = e^{-\frac{(x-E_0)^2}{2\Delta^2}}.
\end{align}
Our motivation is twofold: (1) taking a microcanonical ensemble is a more precise measure of the local eigenvalue statistics, playing the role of ``unfolding'' the spectrum in random matrix theory. (2) As explained in~\cite{SSS}, there is no connected (wormhole) saddle to the bulk equations of motion for $|Z(\beta+iT)|^2$, but there is one for $|Z_{\Delta,N}(T)|^2$.

The spectral form factor $|Z_{\Delta,N}(T)|^2$ has characteristic behavior shown in figure~\ref{fig:SFF}. At early times, the spectral form factor smoothly decays because $Z_{\Delta,N}(T)$ is roughly  the Fourier transform of the window function. When the magnitude of $Z_{\Delta,N}(T)$ becomes sufficiently small, the ramp begins to dominate. The time at which this occurs is called the dip time $T_{\mathrm{dip}}$ and will depend on both $\Delta$ and $N$. The ramp behaves erratically in time with oscillations on a time scale $O(\Delta^{-1})$. However, the time average is linearly growing, which is a smoking gun signature of long range spectral rigidity seen universally in random matrices and chaotic quantum systems. We will show that the dependence of the ramp on $N$ is in a precise sense even more erratic than the dependence on time.

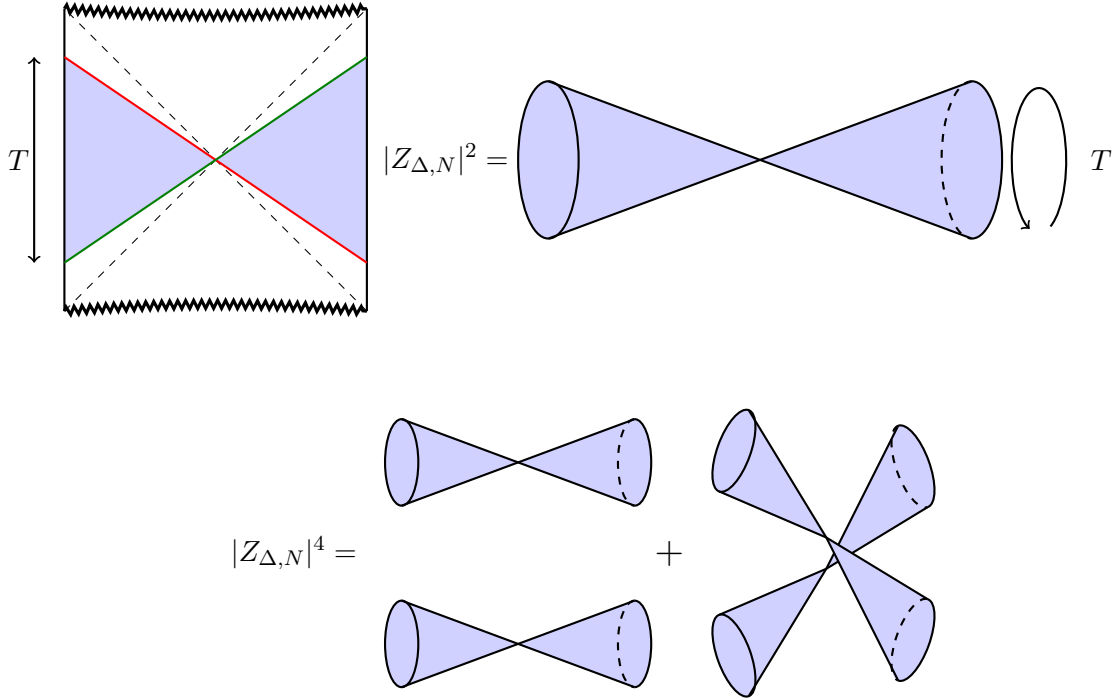
\begin{figure}
    \centering
    \begin{tikzpicture}[scale=.8]

% === LEFT DIAGRAM: Penrose-style diagram ===

% Shaded region bounded by red and green lines (two side bowtie triangles)
\fill[blue!18] (-2.5,4.2) -- (0,2.5) -- (-2.5,0.8) -- cycle;
\fill[blue!18] (2.5,4.2) -- (0,2.5) -- (2.5,0.8) -- cycle;

% Curved zigzag singularities - very subtle inward bow
% Top singularity (bows slightly downward in center)
\draw[decorate, decoration={zigzag, segment length=3pt, amplitude=1.5pt}, very thick]
  (-2.5,5) .. controls (-0.8,4.85) and (0.8,4.85) .. (2.5,5);
% Bottom singularity (bows slightly upward in center)
\draw[decorate, decoration={zigzag, segment length=3pt, amplitude=1.5pt}, very thick]
  (-2.5,0) .. controls (-0.8,0.15) and (0.8,0.15) .. (2.5,0);

% Left and right boundaries
\draw[thick] (-2.5,0) -- (-2.5,5);
\draw[thick] (2.5,0) -- (2.5,5);

% Dashed lines forming X (horizons)
\draw[dashed] (-2.5,0) -- (2.5,5);
\draw[dashed] (2.5,0) -- (-2.5,5);

% Red diagonal line
\draw[red, thick] (-2.5,4.2) -- (2.5,0.8);

% Green diagonal line
\draw[green!50!black, thick] (-2.5,0.8) -- (2.5,4.2);

% T label on right side
\draw[thick, <->] (-3.0,0.8) -- (-3.0,4.2);
\node[right] at (-3.6,2.5) {$T$};

% === RIGHT DIAGRAM: Double cone ===

\begin{scope}[xshift=9cm, yshift=2.5cm]

% Left ellipse fill
\fill[blue!18] (-3.5,0) ellipse (0.5 and 1.3);
% Right ellipse fill
\fill[blue!18] (3.5,0) ellipse (0.5 and 1.3);

% Cone surfaces (filled)
\fill[blue!18] (-3.5,1.3) -- (0,0) -- (-3.5,-1.3) -- cycle;
\fill[blue!18] (3.5,1.3) -- (0,0) -- (3.5,-1.3) -- cycle;

% Cone edges
\draw[thick] (-3.5,1.3) -- (0,0) -- (3.5,1.3);
\draw[thick] (-3.5,-1.3) -- (0,0) -- (3.5,-1.3);

% Left ellipse outline
\draw[thick] (-3.5,0) ellipse (0.5 and 1.3);

% Right ellipse: back half dashed, front half solid
\draw[thick, dashed] (3.5,0) ++(90:1.3) arc[start angle=90, end angle=270, x radius=0.5, y radius=1.3];
\draw[thick] (3.5,0) ++(270:1.3) arc[start angle=270, end angle=450, x radius=0.5, y radius=1.3];

% T circumference arrow using basic arrow tip
\draw[thick, ->] 
  (4.8, -1.1) arc[start angle=-65, end angle=250, x radius=0.45, y radius=1.2];
\node[right] at (5.3, 0) {$T$};

    \end{scope}

    % === TWO STACKED DOUBLE CONES (smaller, no arrow, no T) ===

    % Upper double cone
    \begin{scope}[xshift=5cm, yshift=-2.5cm, scale=0.55]
    \fill[blue!18] (-3.5,0) ellipse (0.5 and 1.3);
    \fill[blue!18] (3.5,0) ellipse (0.5 and 1.3);
    \fill[blue!18] (-3.5,1.3) -- (0,0) -- (-3.5,-1.3) -- cycle;
    \fill[blue!18] (3.5,1.3) -- (0,0) -- (3.5,-1.3) -- cycle;
    \draw[thick] (-3.5,1.3) -- (0,0) -- (3.5,1.3);
    \draw[thick] (-3.5,-1.3) -- (0,0) -- (3.5,-1.3);
    \draw[thick] (-3.5,0) ellipse (0.5 and 1.3);
    \draw[thick, dashed] (3.5,0) ++(90:1.3) arc[start angle=90, end angle=270, x radius=0.5, y radius=1.3];
    \draw[thick] (3.5,0) ++(270:1.3) arc[start angle=270, end angle=450, x radius=0.5, y radius=1.3];
    \end{scope}

    % Lower double cone
    \begin{scope}[xshift=5cm, yshift=-5.5cm, scale=0.55]
    \fill[blue!18] (-3.5,0) ellipse (0.5 and 1.3);
    \fill[blue!18] (3.5,0) ellipse (0.5 and 1.3);
    \fill[blue!18] (-3.5,1.3) -- (0,0) -- (-3.5,-1.3) -- cycle;
    \fill[blue!18] (3.5,1.3) -- (0,0) -- (3.5,-1.3) -- cycle;
    \draw[thick] (-3.5,1.3) -- (0,0) -- (3.5,1.3);
    \draw[thick] (-3.5,-1.3) -- (0,0) -- (3.5,-1.3);
    \draw[thick] (-3.5,0) ellipse (0.5 and 1.3);
    \draw[thick, dashed] (3.5,0) ++(90:1.3) arc[start angle=90, end angle=270, x radius=0.5, y radius=1.3];
    \draw[thick] (3.5,0) ++(270:1.3) arc[start angle=270, end angle=450, x radius=0.5, y radius=1.3];
    \end{scope}

    % === EQUATION LABELS AND CONNECTED CONTRIBUTION ===

    % Y label for the top row (to the left of double cone)
    \node at (3.8, 2.5) {$|Z_{\Delta,N}|^2 =$};

    % Y^2 label for the bottom row
    \node[left] at (2.5, -4) {$|Z_{\Delta,N}|^4 =$};

    % + sign between disconnected and connected contributions
    \node at (7.5, -4) {\Large $+$};

    % Connected contribution: two horizontal double cones with offset tips
    \begin{scope}[xshift=10cm, yshift=-4cm, scale=0.55]
    % Cone A (shifted slightly up)
    \begin{scope}[rotate=20, yshift=-2cm]
    \fill[blue!18] (-3.5,0) ellipse (0.5 and 1.3);
    \fill[blue!18] (3.5,0+3.5) ellipse (0.5 and 1.3);
    \fill[blue!18] (-3.5,1.3) -- (0,0+1.5) -- (-3.5,-1.3) -- cycle;
    \fill[blue!18] (3.5,1.3+3.5) -- (0,0+1.5) -- (3.5,-1.3+3.5) -- cycle;
    \draw[thick] (-3.5,1.3) -- (0,0+1.5) -- (3.5,1.3+3.5);
    \draw[thick] (-3.5,-1.3) -- (0,0+1.5) -- (3.5,-1.3+3.5);
    \draw[thick] (-3.5,0) ellipse (0.5 and 1.3);
    \draw[thick, dashed] (3.5,0+3.5) ++(90:1.3) arc[start angle=90, end angle=270, x radius=0.5, y radius=1.3];
    \draw[thick] (3.5,0+3.5) ++(270:1.3) arc[start angle=270, end angle=450, x radius=0.5, y radius=1.3];
    \end{scope}
    % Cone B (shifted slightly down)
    \begin{scope}[rotate=-20, yshift=2cm]
    \fill[blue!18] (-3.5,0) ellipse (0.5 and 1.3);
    \fill[blue!18] (3.5,-3.5) ellipse (0.5 and 1.3);
    \fill[blue!18] (-3.5,1.3) -- (0,0-1.5) -- (-3.5,-1.3) -- cycle;
    \fill[blue!18] (3.5,1.3-3.5) -- (0,0-1.5) -- (3.5,-1.3-3.5) -- cycle;
    \draw[thick] (-3.5,1.3) -- (0,0-1.5) -- (3.5,1.3-3.5);
    \draw[thick] (-3.5,-1.3) -- (0,0-1.5) -- (3.5,-1.3-3.5);
    \draw[thick] (-3.5,0) ellipse (0.5 and 1.3);
    \draw[thick, dashed] (3.5,0-3.5) ++(90:1.3) arc[start angle=90, end angle=270, x radius=0.5, y radius=1.3];
    \draw[thick] (3.5,0-3.5) ++(270:1.3) arc[start angle=270, end angle=450, x radius=0.5, y radius=1.3];
    \end{scope}
    \end{scope}
\end{tikzpicture}
\caption{Top: the double cone wormhole is constructed by identifying the green and red lines separated by Killing time $T$. Bottom: in the $2n$-boundary computation of $|Z_{\Delta,N}|^{2n}$, there is a $n!$-fold degeneracy of double cone saddles. We have shown the case of $n = 2$.}
    \label{fig:doublecone}
\end{figure}

This linear ramp was precisely reproduced by the so-called double cone wormhole~\cite{SSS}, a universal and (mostly) Lorentzian wormhole constructed by periodically identifying the maximally extended AdS black hole in Killing time $T$ (see figure~\ref{fig:doublecone}). This saddle and its perturbative corrections smoothly depend on $G_{\mathrm{N}}$. Moreover, if one sets up asymptotic boundary conditions for $Z_{\Delta,N}(T)^n\overline{Z_{\Delta,N}(T)}{}^m$ for integers $n$ and $m$, if $n\neq m$, it is not possible to completely pair up the factors of $Z$ and $\bar Z$ via double cones, while if $n=m$, there are $n!$ degenerate ways to do so. This means that $Z_{\Delta,N}(T) $ has the statistics of a complex Gaussian random variable
which is also  the behavior of $Z_{\Delta,N}(T)$ that is found in random matrix theory.

Typically, statistical behavior only emerges  when a quantity of interest is drawn from a random ensemble of some sort. But in the context of AdS/CFT duality, it is far from clear what a suitable ensemble might be, as typical instances of this duality depend on very few parameters, in some cases only a single integer $N$.   We  view the double cone wormhole at times after $T_{\mathrm{dip}}$ and its relation to the microcanonical spectral form factor in the dual CFT as a particularly  sharp example of the factorization puzzle, because the behavior of both sides of the duality are fairly well understood.\footnote{Of course, the spectral form factor has not been explicitly computed in holographic CFT's. Nevertheless, the erratic behavior in time is a general feature of quantum chaos and random matrix theory.} For explicitness, we will focus in this section on the duality between $\mathcal{N} = 4$ super Yang Mills theory and type IIB string theory on $\mathrm{AdS}_5 \times \mathrm{S}^5$, though analogous statements can be made for other theories. 

\subsection{$N$ correlations in the double cone}
\label{sec:doublecone}
In order to analyze to what extent Mellin averaging can account for wormholes, we must understand how theories with different values of $N$ are correlated. There is a bulk method of computing this correlation in AdS/CFT.\footnote{We are indebted to Juan Maldacena for explanations of this computation.} Consider the following boundary observable
\begin{align}
    Z_{\Delta,N}(T)\overline{Z_{\Delta,N+\delta N}(T)}
\end{align}
where $\delta N$ is an integer much less than $N$. 
Considerations similar to the following have previously appeared in~\cite{Cotler:2022rud, Mahajan:2021maz}.

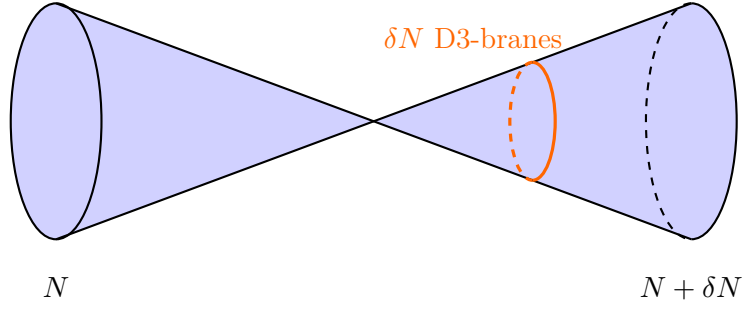
\begin{figure}
    \centering
\begin{tikzpicture}[scale=1.2]

% Left ellipse fill
\fill[blue!18] (-3.5,0) ellipse (0.5 and 1.3);
% Right ellipse fill
\fill[blue!18] (3.5,0) ellipse (0.5 and 1.3);

% Cone surfaces (filled)
\fill[blue!18] (-3.5,1.3) -- (0,0) -- (-3.5,-1.3) -- cycle;
\fill[blue!18] (3.5,1.3) -- (0,0) -- (3.5,-1.3) -- cycle;

% Cone edges
\draw[thick] (-3.5,1.3) -- (0,0) -- (3.5,1.3);
\draw[thick] (-3.5,-1.3) -- (0,0) -- (3.5,-1.3);

% Left ellipse outline
\draw[thick] (-3.5,0) ellipse (0.5 and 1.3);

% Right ellipse: back half dashed, front half solid
\draw[thick, dashed] (3.5,0) ++(90:1.3) arc[start angle=90, end angle=270, x radius=0.5, y radius=1.3];
\draw[thick] (3.5,0) ++(270:1.3) arc[start angle=270, end angle=450, x radius=0.5, y radius=1.3];

% Middle ellipse on the right cone (at about 1/3 from tip to right end)
% At x=1.75, the cone radius scales to 1.3*(1.75/3.5) = 0.65
% x-radius scales similarly: 0.5*(1.75/3.5) = 0.25
% Back half dashed, front half solid
\draw[orange!80!red, very thick, dashed] (1.75,0) ++(90:0.65) arc[start angle=90, end angle=270, x radius=0.25, y radius=0.65];
\draw[orange!80!red, very thick] (1.75,0) ++(270:0.65) arc[start angle=270, end angle=450, x radius=0.25, y radius=0.65];

% Labels
\node[below] at (-3.5,-1.6) {$N$};
\node[below] at (3.5,-1.6) {$N + \delta N$};
\node[above right, orange!80!red] at (0,0.7) {$\delta N$ D3-branes};

\end{tikzpicture}
\caption{The double cone wormhole interpolating between theories with different values of $N$. $\delta N$ probe D3-branes extremize their action by finding a constant, complex radial coordinate to sit at.}
    \label{fig:doubleconedbrane}
\end{figure}

The gravitational path integral leads to a connected contribution, i.e.~a contribution not present in the product of the bulk computations of $Z_{\Delta,N}$ and $\overline{Z_{\Delta,N+\delta N}}$ separately. 
This is a double cone geometry that interpolates between boundary theories with gauge group $U(N)$ and $U(N+\delta N)$. To account for this mismatch, one includes $\delta N$ D3-branes wrapping the spatial $S^3$ and periodic time, but at fixed radial position. When crossing these D3-branes, the five-form flux jumps by $\delta N$  (see figure~\ref{fig:doubleconedbrane}). These branes have tension of $O(N)$ which is much less than the $O(N^2)$ tension needed for backreaction at leading order. We can thus treat them as probe branes and evaluate their contribution to the double cone path integral.

The double cone has the same metric as Schwarzschild $\mathrm{AdS}_5$
\begin{align}
    \d s^2 = -f(r) \d t^2 +\frac{\d r^2}{f(r)}+ r^2 \d\Omega_3^2, \quad f(r) = 1+\frac{r^2}{L_{\mathrm{AdS}}^2}-\frac{\mu}{r^2},
\end{align}
with periodic identification in $t \sim t + T$. We set $L_{\mathrm{AdS}}= 1$ from now on. Each  D3 brane has action
\begin{align}
    S_{\mathrm{D3}} = S_{\mathrm{DBI}} + S_{\mathrm{WZ}} = -T_3 \int \d^4 x\sqrt{-h}+T_3\int C_4,
\end{align}
where $T_3$ is the brane tension, $h$ is the induced metric on the world volume, 
and $C_4$ is the RR 4-form
\begin{align}
    C_4 = r^4 \d t \wedge \d\Omega_3.
\end{align}
With a constant $r$ ansatz, and integrating over the $S^3$ and periodic time, one finds
\begin{align}
    S_{\mathrm{D3}} = 2\pi^2 T T_3(-r^3 \sqrt{f(r)}+r^4)= 2\pi^2 T T_3(-x \sqrt{x^2 + x -{\mu}}+x^2)
\end{align}
where $x \equiv r^2$. Varying the action with respect to $x$, we find three saddle points. At large $\mu$, these are 
\begin{align}
    x_1 &= e^{5\pi \i/3}\left(\frac{\mu^2}{2}\right)^{1/3} + e^{\i\pi/3}\left(\frac{\mu}{4}\right)^{1/3} + \dots
    \\
    x_2 &= -\left(\frac{\mu^2}{2}\right)^{1/3} -\left(\frac{\mu}{4}\right)^{1/3} +\dots
    \\
    x_3 &= e^{\i\pi/3}\left(\frac{\mu^2}{2}\right)^{1/3} +  e^{5\pi \i/3}\left(\frac{\mu}{4}\right)^{1/3} + \dots
\end{align}
which are far from the black hole horizon.
When performing the path integral over $x$, we take the original contour to start at large real values of $x$, move inward to the branch point at $x_+$, where $x_+=r_+^2$ is the positive root of $x^2+x-\mu$, and move around it clockwise, bringing the contour to the second side of the double cone and then back out to infinity. 
In order to break the phase degeneracy of the two conjugate saddles, we can multiply the action by $(1-\i\epsilon)$, which is equivalent to giving Newton's constant a small imaginary piece. Upon analysis of the steepest ascent and descent lines, we find that only the saddle at $x_1$ contributes to the integral.
The real part of $\i S$ at this saddle is
\begin{align}
    \Re[iS(x_1)]=-TT_3\frac{3 \sqrt{3}\pi^2  \mu ^{2/3}}{4 \sqrt[3]{2}} + \dots
    \label{eq:reis}
\end{align}
Because $\delta N \ll N$, the collective contribution to the path integral is 
\begin{align}
    \la Z_{\Delta,N}(T)\overline{Z_{\Delta,N+\delta N}(T)}\ra_c \sim T e^{-\# T N \delta N },
    \label{eq:1.10}
\end{align}
where $\#$ is an $O(1)$ number from \eqref{eq:reis}. The multiplicative factor of $T$ is from the zero mode of the relative time shift of the two boundaries of the double cone while the exponent comes from the action of the complex saddle point of the D3-brane. There is also a rapidly oscillating phase, though this is not relevant when considering the spectral form factor itself.

We can interpret this answer as giving a correlation length in $\delta N$
\begin{align}
    \xi_{\delta N} \equiv \frac{1}{\# N T}.
\end{align}
This means that if $\delta N$ is larger than $\xi_{\delta N}$, the spectral form factors are effectively independent variables. Because $\xi_{\delta N} \ll 1$, the spectral form factor at consecutive integer values of $N$ are independent. If we assume that this result makes sense for real values of $N$, then there are roughly $N T$ independent spectral form factors between $N$ and $N + 1$. This suggests that at superpolynomial times (that is, for $T$ greater than any power of $N$), there are enough independent samples for Mellin averaging to reproduce the full perturbative expansion around the double cone including its Gaussian statistics. In particular, the statistical error of averaging over an $O(1)$ window in $N$ is $O(1/\sqrt{NT})$, which is subdominant to the term of order $N^{-K}$ for all $K$ if $T$ is superpolynomially large.

In Mellin averaging, we consider a fixed function of $N$, so once we start scaling $T$ with $N$, e.g.~$T = N^p \tilde{T}$, the relevant correlation is between $Z_{\Delta,N}(N^p \tilde{T})$ and $Z_{\Delta,N+\delta N}((N+\delta N)^p \tilde{T})$.\footnote{We thank Yiming Chen for emphasizing this point.} There is no known saddle giving a connected contribution to this observable because the times on the two sides of the wormhole are very different. This suppresses the correlations even further, as we argue at the end of section~\ref{sec:loop}, though we expect this does not qualitatively change the conclusion that superpolynomial times are needed for Mellin averaging to successfully account for wormholes.

\subsection{The dip time}

Because the correlation length decays in time, it is critical to understand what is $T_{\mathrm{dip}}$, where the double cone dominates but the fluctuations in $N$ are largest. 

In a fixed theory, the spectral density is a sum of delta functions. However, at energy scales much larger than the level spacing, for some questions, this sum of delta functions can be approximated as a smooth density of states, $\rho(E)$, with perhaps a non-analyticity at the spectral edge. 
If this assumption is valid for the computation of the spectral form factor for all times before the level repulsion of the linear ramp kicks in and we take a small enough microcanonical window such that $\rho(E)$ is roughly constant, we would have that the spectral form factor is approximately given by the Fourier transform of the window function
\begin{align}
   |Z_{\Delta,N}(T)|^2 \approx  \rho(E_0)^2\left|\int \d x e^{-\i x T}  e^{-\frac{(x-E_0)^2}{2\Delta^2}}\right|^2 \sim e^{2S(E_0)-\Delta^2 T^2}.
\end{align}
Such an assumption is valid in random matrix theory.
Equating this with the linear ramp, we find a dip time of $\sim \tfrac{\sqrt{S(E_0)}}{\Delta}$, which is $O(N)$ if we take $\Delta$ of $O(1)$. This is a relatively short time and so would seem to invalidate Mellin averaging.

We will argue that the smooth approximation to $\rho(E)$ is invalid in holographic conformal field theories at the dip time because of atypical states in the spectrum. Even if the entropy of these states is small compared to the black hole entropy, they can, in principle, delay the dip time by a large, even superpolynomial, amount.

\begin{figure}
    \centering
    \includegraphics[width=\textwidth]{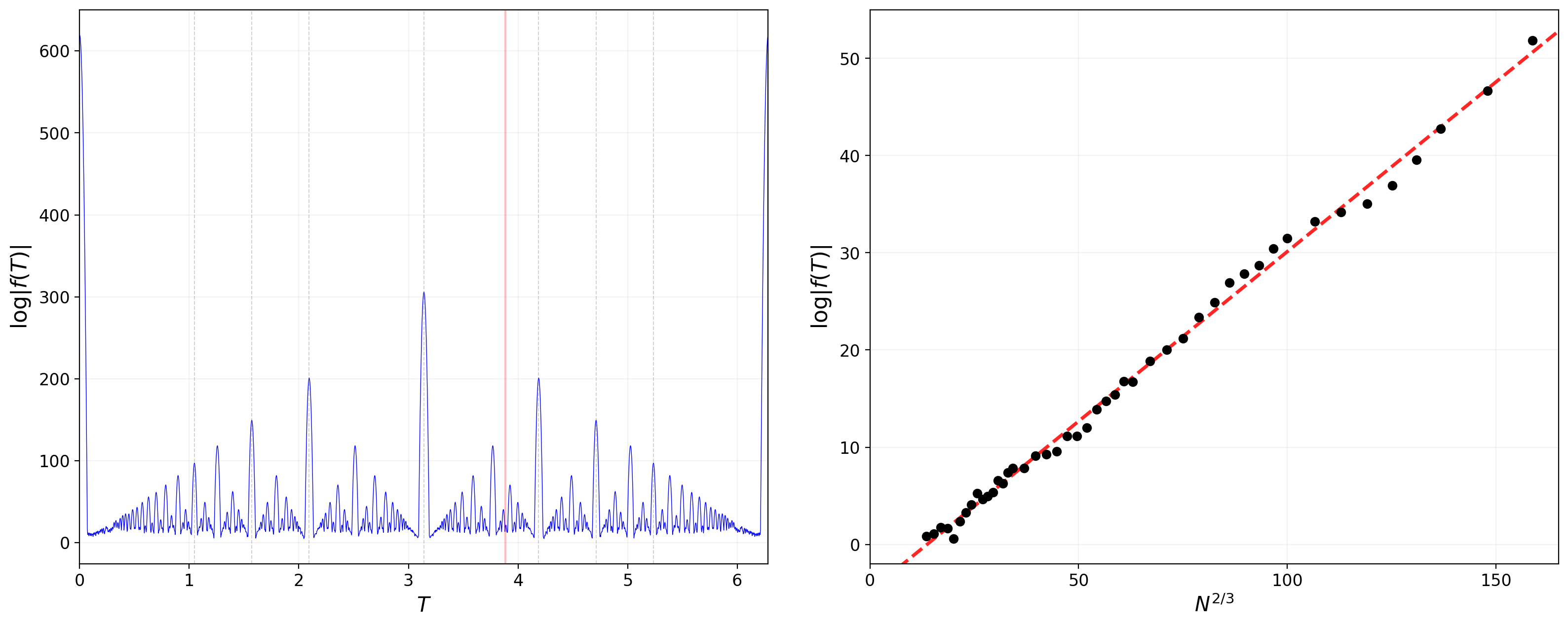}
    \caption{Left: the BPS spectral form factor for $U(500)$ and  $\Delta = N$, and $E_0 = \tfrac{N^2}{4}$. Right: the value of the BPS spectral form factor at $T=2\pi/\phi$ where $\phi$ is the golden ratio marked as a red vertical line on the left.}
    \label{fig:BPSSFF}
\end{figure}

As an illustrative example, we can consider the contribution from $1/2$ BPS states, which are numerous but extremely atypical. $1/2$ BPS states with energy $n$ correspond to operators of the form
\begin{align}
    \Tr(Z^{a_1})\Tr(Z^{a_2})\dots \Tr(Z^{a_j}), \quad \sum_i a_i=n
\end{align}
where $Z = \Phi_1 + \i \Phi_2$ is a chiral primary field. Their spectrum is exactly integer spaced but with large degeneracies, $d(n)$. This leads to phase coherence in the spectral form factor that is not captured by the smooth approximation of $\rho(E)$. The contribution of  this subsector to $Z_{\Delta,N}(T)$ is
\begin{align}    f(T) \equiv \Tr_{\mathrm{BPS}}(e^{-\i H_NT }w_\Delta(H_N)) 
    = \sum_{n}d(n)e^{-(n-E_0)^2/2\Delta^2}e^{-\i Tn}
\end{align}
where $d(n)$ is the degeneracy and $\Tr_{\mathrm{BPS}}$ is the trace only over this subsector.
We note that $f(T)$ is $2\pi $ periodic, so we can focus on $T\in[0,2\pi)$. 

The generating function for the degeneracy when the gauge group is $U(N)$ is
\begin{align}
    \mathcal{Z}(z) \equiv \sum_{n = 0}^\infty
d(n)z^n = \prod_{k=1}^{N}\frac{1}{1-z^k}
\end{align}
so the degeneracy is extracted as
\begin{align}
    d(n) = \frac{1}{2\pi \i}\oint \frac{1}{z^{n+1}}\prod_{k=1}^{N}\frac{1}{1-z^k}\d z
\end{align}
where the contour wraps around $z = 0$ counterclockwise once.

The condition for an extremum of the integrand is
\begin{align}
    -\frac{n+1}{z}+\sum_{k = 1}^N\frac{k z^{k-1} }{1-z^k} = 0
\end{align}
which is equivalent to a high degree polynomial equation, so there are many solutions. In particular, for any positive integer $q \ll N$, there is a saddle point near $z = \omega$, where $\omega$ is a primitive $q^{th}$ root of unity. The statement that the saddle point is ``near'' $z = \omega$ means that keeping $q$ and $n/N^2$ fixed, the saddle approaches $z = \omega$ as $N \rightarrow \infty$. The saddle point that dominates $d(n)$ is the one near $z = 1$. However, the contribution from this saddle point to $|f(T)|$ decays rapidly away from $T =0 $ because the contribution to $d(n)$ of the dominant saddle point is an analytic function that is not rapidly varying in $n$. To find the dominant behaviour of $|f(T)|^2$ near $T = 2\pi p/q$ where $p$ and $q$ are relatively prime, we instead need to look at a saddle point near $z =\omega$ where $\omega$ is a primitive $q^{th}$ root of unity.

With this in mind, we set $z =\omega e^{-s/N}$ with $s\ll N$. Using $\prod_{j=0}^{q-1}(1-\omega^j y)=1-y^q$ on blocks of $q$ consecutive values of $k$, we find that
\begin{align}
    \mathcal{Z}(\omega e^{-s/N}) \approx \mathcal{Z}(e^{-qs/N})^{1/q}.
\end{align}
The saddle point is therefore at $s=u/q$, where, writing $k=xN$ in the extremum condition at $q=1$, $u$ is fixed by $\int_0^1 \d x \frac{x}{e^{ux}-1}=\frac{n}{N^2}$, and its exponent is $1/q$ times that of the saddle near $z=1$, which is $\approx \frac{N}{q}\left(\log \frac{n}{N^2}+2\right)$ for $n\gg N^2$. This saddle contributes $\omega^{-n}A(n)$ to $d(n)$ with $A(n)$ slowly varying, so it makes a large contribution to $f(T)$ only if $T$ is close to $2\pi \frac{p}{q}$.
This contribution is of order $\sim e^{\frac{N}{q}}$, where we have dropped subleading multiplicative factors. The Gaussian microcanonical window leads to a Gaussian decay about each saddle of width $\Delta^{-1}$.

Now that we have determined that there is a resonance at rational times $2\pi p/q$ of order $e^{N/q}$, we can investigate the floor of the function at irrational values.  If $\Delta \ll  \sqrt{N}$, then the saddles at $T = 2\pi \mathbb{Z}$ will dominate for all values of $T$ and so the floor will be $\sim e^N$. For larger $\Delta$, we must consider the effect of other saddles. For any irrational number $T/2\pi$, Hurwitz's theorem in the study of Diophantine approximation says that there are an infinite number of relatively prime pairs $p/q$ with 
\begin{align}
    \left| \frac{T}{2\pi} - \frac{p}{q}\right|< \frac{1}{\sqrt{5}q^2}
\end{align}
where the $\sqrt{5} $ is optimal.
Together with the saddle point value and Gaussian decay, this suggests an effective action for contribution of saddle points near primitive $q^{th}$ roots of 1 to the floor value of the $1/2$ BPS spectral form factor 
\begin{align}
    S_{\mathrm{eff}}[q] \sim c_1\frac{N}{q} - c_2\frac{ \Delta^2}{q^4}
\end{align}
where $c_{1/2}$ are $O(1)$. The maximum of this function is at $q_* \sim (\tfrac{\Delta^2}{N})^{1/3}$, giving a floor height of $\sim e^{\#\frac{N^{4/3}}{\Delta^{2/3}}}$, which is superpolynomial. This scaling is numerically verified in figure~\ref{fig:BPSSFF} for $\frac{T}{2\pi}$ equal to the inverse of the golden ratio, which is in a sense maximally irrational.

With this superpolynomial floor, the double cone geometry can only dominate the gravitational path integral at superpolynomial times. At these times, Mellin averaging is potentially a viable resolution to the factorization puzzle.

However, in highlighting the importance of the $1/2$ BPS spectrum in resolving the factorization puzzle, we have set up a straw man because these states necessarily have large R-charge. Instead of looking at the standard spectral form factor, we could instead ask about the charge resolved spectral form factor. For example, if we go to the zero charge sector, the double cone saddle would still contribute and the  $1/2$ BPS states would not. We should similarly focus on the zero angular momentum sector. A sufficient condition on the spectrum for subdominance of the double cone would be to have a class of neutral states at high energies that have a non-chaotic spectrum, giving a superpolynomial floor. These could be extremely atypical and would not need to be exact eigenstates, only to have sufficiently long lifetimes. Interesting non-examples include dilute gases of high angular momentum gravitons with total angular momentum adding to zero and ``oscillons''~\cite{Maliborski:2013jca,Milekhin:2023was}, stable periodic solutions of scalar fields in AdS. The former are  insufficient because perturbative gravitational interactions will lead to their collapse in polynomial time. The latter are insufficient because they have a maximum energy (of order $N^2$) before collapse. The best we can say currently is that it is difficult to prove that sufficient neutral states do not exist in the spectrum.

It would be valuable to understand other cases in which wormholes dominate the path integral. At present, the double cone is the only case we are aware of where the wormhole is understood to dominate and both the bulk and boundary sides of the computation are relatively well understood. In these other cases, it is not a priori clear that atypical states will be relevant.

\section{The Spectral Form Factor in Random Matrix Theory}
\label{sec:loop}

In order to gain intuition on the formula \eqref{eq:1.10} for correlations between different values of $N$, and whether this formula makes sense for  real values of $N$, we will analyze a toy model of holographic CFTs based on random matrices. We consider a fixed $L \times L$ matrix $H$ drawn from an invariant ensemble with potential $V(H)$. Here, $L$ is the Hilbert space dimension related to the number of qubits as $L = 2^N$. We can model the removal of a single qubit by projecting $H$ to an $L/2 = 2^{N-1}$ dimensional subspace, $PHP$, where $P$ is a rank $L/2$ projector. This can be thought of as setting $L/2$ of the rows and columns of $H$ to zero. $H$ and $PHP$ are clearly correlated because they share $L^2/4$ of the same matrix elements. Instead of removing a full qubit, we can remove a fractional number of qubits by taking the rank of $P$, $\alpha L$, to be larger than $L/2$ i.e.~setting fewer rows and columns to zero. In the $L\rightarrow\infty$ limit, this defines a family of correlated theories that depend on a real value of qubits, $\log_2 \alpha L$.

Let us work with potential $V(H)$ for one-cut matrix models, with support $[a,b]$ for $L\times L$ matrices. We define the resolvents
\begin{align}
    R_H(x) = \Tr\left(\frac{1}{x -H}\right), \quad R_{PHP}(y) = \Tr\left(P\frac{1}{y -PHP}P\right),\quad \Tr(P) = \alpha L.
\end{align}
We will compute the connected correlator of the resolvents, $\la R_H(x) R_{PHP}(y)\ra_c$, which in turn will lead to the connected correlator of spectral form factors.

For the Gaussian Unitary Ensemble (GUE), we can solve for this using combinatorial tools (appendix~\ref{sec:combinatorics}), but for more general matrix models, we use loop equations. A useful loop equation for this correlator is
\begin{align}
\begin{aligned}
     0 = \sum_{ab} \int \d H \frac{\partial}{\partial H_{ab}}\left[(x-H)^{-1}_{ab} R_{PHP}(y) e^{-L\Tr V(H)} \right].
\end{aligned}
\end{align}
By considering the derivative on each of the three terms, we find that
\begin{align}
\begin{aligned}
    0 = \la R_H(x) R_H(x) R_{PHP}(y)\ra + \la \Tr((x-H)^{-1}P(y-PHP)^{-2}P)\ra-
    \\ L \la \Tr((x-H)^{-1}V'(H))R_{PHP}(y)\ra.\label{eq:loop1}
\end{aligned}
\end{align}

We also have the standard loop equation~\cite{EynardOrantin:2007}
\begin{align}
    0 = \sum_{ab} \int \d H \frac{\partial}{\partial H_{ab}}\left[(x-H)^{-1}_{ab}  e^{-L\Tr V(H)} \right].
\end{align}
which leads to 
\begin{align}
    0 = \la R_H(x)R_H(x)\ra- L\la \Tr((x-H)^{-1}V'(H))\ra.
\end{align}
Subtracting this equation (times $\la R_{PHP}(y)\ra$) from \eqref{eq:loop1} gives
\begin{align}
\begin{aligned}
         \la R_H(x)R_H(x) R_{PHP}(y)\ra_c+2\la R_H(x)\ra\la R_H(x)R_{PHP}(y)\ra_c+ \la \Tr((x-H)^{-1}P(y-PHP)^{-2}P)\ra
    \\
    =L \la \Tr((x-H)^{-1}V'(H))R_{PHP}(y)\ra_c
\end{aligned}
\label{eq:loop2}
\end{align}
Trivially, the right hand side equals
\begin{align}
\begin{aligned}
       \la \Tr((x-H)^{-1}V'(H))R_{PHP}(y)\ra_c = V'(x)\la R_H(x)R_{PHP}(y)\ra_c 
    \\-\la \Tr((x-H)^{-1}(V'(x)-V'(H)))R_{PHP}(y)\ra_c 
\end{aligned}
\end{align}
Under the assumption that $V(x)$ is a polynomial, $\Tr((x-H)^{-1}(V'(x)-V'(H)))$ is also a polynomial in $x$ with coefficients that are integer moments of $H$. Thus, the entire second term on the RHS of the above equation must be polynomial in $x$, and so does not have any singularities in $x$. Therefore,
\begin{align}
\begin{aligned}
           2\la R_H(x)\ra\la R_H(x)R_{PHP}(y)\ra_c + \la R_H(x)R_H(x)R_{PHP}(y)\ra_c+ \la \Tr((x-H)^{-1}P(y-PHP)^{-2}P)\ra
        \\
    =L V'(x)\la R_H(x)R_{PHP}(y)\ra_c + L\mathrm{Poly}_{\alpha}(x;y).
\end{aligned}\label{longidentity}
\end{align}
We now take the large $L$ limit, defining the $O(L^0)$ quantities
\begin{align}
    G_H(x) &\equiv \lim_{L\rightarrow \infty} \frac{1}{L}\la R_H(x)\ra
    \\
    G_{PHP}(x) &\equiv \lim_{L\rightarrow \infty} \frac{1}{\alpha L}\la R_{PHP}(x)\ra\label{eq:rp}
    \\
    s_\alpha(x,y) &\equiv \lim_{L\rightarrow \infty} \frac{1}{L}\la 
     \Tr((x-H)^{-1}P(y-PHP)^{-1}P)\ra
    \\
    W_\alpha(x,y) &\equiv \lim_{L\rightarrow \infty} \la R_H(x) R_{PHP}(y)\ra_c
\end{align}
With this convention, $G_{PHP}(x) \sim 1/x$ at large $x$. 
The leading order equation in $L$ is then
\begin{align}
         -\partial_ys_\alpha(x,y)
    = (V'(x)-2G_H(x))W_\alpha(x,y) + \mathrm{Poly}_{\alpha}(x;y)  \label{loopeq}
\end{align}
and $\mathrm{Poly}_{\alpha}(x;y)$ is uniquely fixed by requiring that $W_\alpha(x,y) = O(x^{-2})$ at large $x$ and similarly for $y$, and that $W_\alpha$ has no singularities away from the cut.  In deriving this equation, we drop the term $ \la R_H(x) R_H(x) R_{PHP}(y)\ra_c$ in \eqref{longidentity} as it is $O(1/L)$,
while other terms in the equation are of order $L$.

$H$ and $P$ are freely independent variables, which essentially means that their eigenbases are Haar randomized with respect to each other.
That is so because $P$ is a fixed projection operator but $H$ is drawn from an ensemble with full unitary symmetry, so its eigenvectors are
randomized with respect to those of $P$.  Free independence is the natural analog for random matrices of statistical independence for ordinary random variables.   Just as statistical independence lets
one compute joint moments of a product in terms of moments of the individual variables, free independence of $H$ and $P$ lets one compute the expectation value of the trace of a function of $H$
and $P$, such as $s_\alpha(x,y)$, in terms of expectation values of functions of one
of the two variables only.   For a simple function, the procedure is simple.   For example, averaging over $H\to U H U^{-1}$ for $U\in U(L)$ will, in a standard way, enable one to express
$\frac{1}{L}\la\Tr\,H P H P\ra$ in terms of $\frac{1}{L}\la \Tr\,H^2\ra$ as well as
$\frac{1}{L}\la \Tr \,P\ra=\alpha$.

For a more complicated function such as $s_\alpha(x,y)$, one has to use machinery of free probability theory. 
  First, observe that free independence of $H$ and $P$ is a reciprocal relation;
instead of keeping $P$ fixed and averaging $H$ over unitary conjugation, we could equivalently keep
$H$ fixed and average $P$ over unitary conjugation.   This gives a map from functions of $H$ and $P$
to functions of $H$ only.   That map is called a conditional expectation (from the algebra generated by $H$ and $P$ to the algebra generated by $H$ only).   We denote the conditional expectation as $E[~\cdot~]$.   As shown in appendix~\ref{sec:proof}, one has in particular 
\be\label{oddrel} E\left[P(y-PHP)^{-1}P\right]=(\omega(y)-H)^{-1},\ee where $\omega(y)$ is defined implicitly by
\begin{align}
    {\omega(y) = y+\frac{1-\alpha}{G_H(\omega(y))}}
    \label{eq:wdef}.\end{align}
So (using $\frac{1}{AB}=(\frac{1}{A}-\frac{1}{B})\frac{1}{B-A}$)
\begin{align}
\label{eq:s0}
    {s_\alpha(x,y)= \lim_{L\to\infty}\frac{1}{L}\la\Tr\,(x-H)^{-1}(\omega(y)-H)^{-1}\ra =\frac{  G_H(x)-G_H(\omega(y))}{\omega(y)-x}}.
\end{align}
Taking the trace of \eqref{oddrel}, we learn also that the normalized resolvents of $H$ and $PHP$ are related by 
a reparametrization of the spectrum:
\be\label{oddchange} G_{PHP}(y)=\frac{1}{\alpha}G_H(\omega(y)).\ee

Substituting the result \eqref{eq:s0} for $s_\alpha(x,y)$ in the loop equation \eqref{loopeq}, we get
\be\label{substitloop}   -\partial_y\left(\frac{  G_H(x)-G_H(\omega(y))}{\omega(y)-x}\right)
    = (V'(x)-2G_H(x))W_\alpha(x,y) + \mathrm{Poly}_{\alpha}(x;y)   \ee Using $\partial_y=(\partial_y\omega(y))\partial_\omega$, the left hand side is
    \be\label{subtwo} -\partial_y\omega(y)\cdot\left.\partial_\omega \left(\frac{  G_H(x)-G_H(\omega)}{\omega-x}\right) \right|_{\omega=\omega(y)} .\ee 
Specializing \eqref{substitloop} to $\alpha=1$, and using the fact that $\omega(y)=y$ at $\alpha=1$, we have 
\be\label{alphaone}  -\partial_y\left(\frac{  G_H(x)-G_H(y)}{y-x}\right)
    = (V'(x)-2G_H(x))W_1(x,y) + \mathrm{Poly}_1(x;y)  . \ee   
    Comparing these last formulas, we can now recognize that
\begin{align}
    W_\alpha(x,y) = W_1(x,\omega(y))\partial_y \omega(y).
    \label{eq:Wa}
\end{align} and similarly that $\mathrm{Poly}_{\alpha}(x;y)=\partial_y\omega(y)\cdot 
\mathrm{Poly}_1(x;\omega(y))$.

{At $\alpha=1$, $P=1$, so $W_1(x,y)$ reduces to an observable of a one-matrix
 model, namely $\la R_H(x) R_H(y)\ra_c$.
This  is known to be universal in the sense that it only depends on the  endpoints of the eigenvalue distribution, not on further details of the matrix potential.}  
For  the symmetric case of endpoints $[-a,a]$, one has
\begin{align}\label{onecut}
    W_1(x,y) = \frac{1}{2(x-y)^2}\left( \frac{xy-a^2}{\sqrt{(x^2-a^2)(y^2-a^2)}}-1\right).
\end{align}
Thus, the knowledge of $G_H(x)$ fixes $W_\alpha(x,y)$.

As an illustrative example, we consider the Gaussian unitary ensemble, with $a =2$. A quartic interaction is considered in appendix~\ref{sec:quartic}. The single matrix resolvent for the GUE is well-known~\cite{mingo2017free}:
\begin{align}
    G_H(x) = \frac{x-\sqrt{x^2-4}}{2}.
\end{align}
Solving \eqref{eq:wdef}, we find
\begin{align}
   {\omega(y)=\frac{(\alpha +1) y-(\alpha -1) \sqrt{y^2-4 \alpha }}{2 \alpha }}.
\end{align}
It is useful to uniformize the coordinates as 
\begin{align}
    x(z) = \left( z+\frac{1}{z}\right),
    \quad
    y(\zeta) =  \sqrt{\alpha}\left( \zeta+\frac{1}{\zeta}\right).
\end{align}
$z$ and $\zeta$ are single-valued on the complex plane. The physical sheet is mapped to the exterior of the unit disc and the second sheet is mapped to the interior.
This uniformization leads to simple formulas such as
\be\label{simpleones} \omega(y(\zeta))=\frac{\zeta}{\sqrt\alpha}+\frac{\sqrt{\alpha}}{\zeta},~~~~~\frac{\partial\omega}{\partial y}=\frac{1}{\alpha}\frac{\zeta^2-\alpha}{\zeta^2-1}.\ee
Setting $a=2$ in \eqref{onecut} and using \eqref{eq:Wa}, we get 
\begin{align}
    W_\alpha(x(z),y(\zeta))=\frac{\zeta ^2 z^2}{\left(\zeta ^2-1\right) \left(z^2-1\right) \left(\sqrt{\alpha
   }-\zeta  z\right)^2}.
\end{align}

With the goal of eventually
analyzing the spectral form factor, we want to use these results to compute connected
expectations such as $\bigl\la \Tr f(H)\Tr g(PHP)\bigr\ra_c $ for suitable functions $f,g$.   Such an expectation can be computed in terms of the
contour integral  \be\label{contourint} \bigl\la \Tr f(H) \Tr g(PHP)\bigr\ra_c
=\frac{1}{(2\pi i)^2} \oint \,f(x) g(y) W_\alpha(x,y)d x \,d y,\ee with the contours going around
the cuts in the $x$ and $y$ planes.   Such an integral simplifies drastically in the $z,\zeta$
coordinates since $W_\alpha(x,y) d x\, d y=\frac{\sqrt \alpha}{(\sqrt \alpha-\zeta z)^2}d z \, d\zeta$.
So
\be\label{contourint2} \bigl\la \Tr f(H)\Tr g(PHP)\bigr\ra_c
=\frac{1}{(2\pi i)^2} \oint \,f(x) g(y) \frac{\sqrt \alpha}{(\sqrt \alpha-z\zeta)^2}  dz\,d\zeta.\ee
The integral is now taken over the product of  unit circles $|z|=|\zeta|=1$.

It is also very useful to express the functions $f,g$
in a basis that simplifies the integrals.    Such a basis is provided by the Chebyshev polynomials of the first kind.   These are defined by $T_n(\cos\theta)=\cos(n\theta)$, so
\begin{align}
 T_n(\tfrac{x}{2})=   T_n(\tfrac{z}{2}+\tfrac{1}{2z}) = \frac{z^n + z^{-n}}{2}.
\end{align}
  Any function on the interval $[-2,2]$ can be expanded in 
Chebyshev polynomials:
\begin{align}
\begin{aligned}
       f(x) &= a_0 + 2\sum_{n =1}^{\infty}a_n T_n(x/2),
    \quad
    a_n = \frac{1}{\pi}\int_0^{\pi}\d\theta f(2\cos\theta)\cos(n\theta) .
\end{aligned}
\end{align}
{In this basis, the contour integrals are simple and give a diagonal result:}
\begin{align}
\label{eq:chebconn}
\begin{aligned}
\biggl\la \Tr \,T_m(\tfrac{H}{2})\;\Tr\, T_n(\tfrac{PHP}{2\sqrt{\alpha}})\biggr\ra_c =& \frac{1}{(2\pi i)^2}\oint\oint \frac{z^m+z^{-m}}{2} \frac{\zeta^n+\zeta^{-n}}{2}\frac{\sqrt \alpha}{(\sqrt \alpha-z\zeta)^2}d z\,d\zeta
\\
= &\frac{m}{4}\alpha^{m/2}\delta_{nm}.
\end{aligned}
\end{align}

We can now expand the function $ e^{\i T x}w_{\Delta}(x)$ that appears in the definition of
the spectral form factor in this basis. The Chebyshev coefficients are
\begin{align}
    a_n(T) = \frac{1}{\pi} \int_0^{\pi} \d\theta e^{\i 2T\cos\theta}w(2\cos \theta)\cos(n\theta).
\end{align}
so that
\begin{align}
  \biggl  \la \Tr(e^{\i T H}w_{\Delta}(H))\Tr(e^{-\i T \frac{PHP}{\sqrt{\alpha}}}w_{\Delta}(\tfrac{PHP}{\sqrt{\alpha}}))\biggr\ra_c = \sum_{r = 1}^{\infty} r \alpha^{r/2}|a_r(T)|^2.
    \label{eq:sum}
\end{align}
Taking a Gaussian microcanonical window function at the center of the energy spectrum, we compute
\begin{align}
    a_r(T) &= \frac{1}{\pi}\int_0^{\pi} \d\theta e^{\i 2T\cos\theta}e^{-\frac{2\cos^2\theta}{\Delta^2}}\cos (r\theta)
    \\
    &= \frac{1}{\pi}\int_{-\pi/2}^{\pi/2} \d u e^{-\i 2T\sin u}e^{-\frac{2\sin^2u}{\Delta^2}}\cos ( r\pi/2+ru)
\end{align}
We consider $T^{-1/2}\ll \Delta \ll T^{-1/3}$ and $|\log\alpha|\ll (T\Delta^2)^{-1}$. The integral concentrates around small $u$ and we can linearize the $\sin$'s:
\begin{align}
a_r(T)    \approx \frac{1}{\pi}\int_{-\infty}^{\infty} d u e^{-\i 2T  u}e^{-\frac{2u^2}{\Delta^2}}\cos (ru + r\pi/2).
\end{align}
Evaluating the Gaussian integral, we get
\begin{align}
    |a_r(T)|^2 \approx \frac{\Delta ^2 e^{-\frac{1}{4} \Delta ^2 (r-2 T)^2}}{8 \pi }.
\end{align}
Therefore, the sum in \eqref{eq:sum} is dominated by the value of $r$ where the exponent of the summand of \eqref{eq:sum}, $-\frac{1}{4} \Delta ^2 (r-2 T)^2+ \frac{r}{2}\log \alpha$, is maximized, which is $\sim 2T$. We thus find that 
\begin{align}
   \biggl \la \Tr\bigl(e^{\i T H}w_{\Delta}(H)\bigr)\Tr\left(e^{-\i T \frac{PHP}{\sqrt{\alpha}}}w_{\Delta}(\tfrac{PHP}{\sqrt{\alpha}})\right)\biggr\ra_c  \sim  T \alpha^{T}=  T e^{- \delta S~ T }
\end{align}
where in the second equality, we write $\alpha=e^{-\delta S}$, where $\delta S$ is the entropy difference between the two matrices. 
This has the same form as the answer from the D3 brane calculation in $\mathcal{N} = 4$ SYM \eqref{eq:1.10} once we recognize that $\delta S \sim N \delta N$ in this theory because the entropy is quadratic in $N$.

If we instead consider $\bigl\la \Tr(e^{\i T_1 H}w_{\Delta}(H))\Tr(e^{-\i T_2\frac{PHP}{\sqrt{\alpha}}}w_{\Delta}(\frac{PHP}{\sqrt{\alpha}}))\bigr\ra_c$, with $T_1 \neq T_2$, there will be an additional Gaussian suppression of $\sim e^{-\Delta^2(T_1-T_2)^2/4}$. As  mentioned in section~\ref{sec:SFFADSCFT}, this sort of decorrelation is relevant for the Mellin averaging, but does not appear to change the qualitative picture that superpolynomial times are needed. In holography, the double cone is only a saddle when $T_1 = T_2$, but this suppression follows from integrating over double cones with different energies $E$, weighted by $e^{\i E(T_1-T_2)}w_\Delta(E)^2$.\footnote{We thank the referee for this observation.}

\section{Toy Models of Quasirandom Functions}\label{toymodels}

Here we will construct a toy model of a deterministic function on the real axis  whose values at positive integers behave
as independent random draws from a Gaussian ensemble.\footnote{The function we construct simulates draws from a real Gaussian ensemble.  Replacing $\cos(a^N)$ in what follows with $\exp(i a^N)$ gives
a function that similarly simulates draws from a complex Gaussian ensemble.}    This is meant to simulate the sort of thing that is found in studying wormholes. 

A first try is to consider the sequence $F(N)=\cos(a N)$, with $a$ a real constant that is not a rational multiple of $\pi$.  View $aN$ mod $2\pi$ as an angle $\theta_N$.
The angles $\theta_N$ are equidistributed on the unit circle~\cite{Weyl}.   
This statement means that for large $N$, the values of $\theta_1,\theta_2,\dots,\theta_N$  fill up the unit circle uniformly.   Thus, we can view $\theta_N$ for large $N$ as a random
draw from the uniform probability distribution $\frac{\d\theta}{2\pi}$ on the unit circle.   Setting $x=\cos\theta$, this becomes $\frac{\d x}{\pi\sqrt{1-x^2}}$ on the interval $[-1,1]$.   Thus we can view $F(N)=\cos\theta_N$
as a random draw from the probability distribution $\mu=\frac{\d x}{\pi\sqrt{1-x^2}}$ on the interval $[-1,1]$.

However, the $F(N)$ for different values of $N$ are not {\it independent} draws from this probability distribution.   For example, since $F(N+1)=\cos(a(N+1))=\cos a \cos(aN)-\sin a \sin(aN)=\cos a F(N)
\pm\sin a\sqrt{1-F(N)^2}$, the values 
of $F(N)$ and $F(N+1)$ are definitely not independent.   Similarly since $\cos(2aN)=2\cos^2(aN)-1$, $F(N)$ and $F(2N)$ are not independent.

Such relations can be avoided  by replacing $\cos(a N)$ with $\cos( a N^\alpha)$ for a generic positive number $\alpha$.  However, this does not achieve true statistical independence.  For
example, suppose that $0<\alpha<1$.   Then for large $N$, $a (N+1)^\alpha- a N^\alpha\ll 1$, and therefore $\cos(a N^\alpha)$ and $\cos(a (N+1)^\alpha)$ are almost equal -- they are certainly
not statistically independent.   If instead $1<\alpha<2$,  then for large $N$, one has  $(N+2)^\alpha-2(N+1)^\alpha +N^\alpha\ll 1$.  This relation implies that $\cos(a N^\alpha)$, $\cos(a (N+1)^\alpha)$, and
$\cos(a(N+2)^\alpha)$ are not independent: for large $N$, a knowledge of two of these determines a lot about the third.    Something similar happens for any value of $\alpha$.

We can avoid all these issues by defining $F(N)=\cos(a^N)$.   For any transcendental\footnote{An algebraic number is one that is the root of a non-zero polynomial equation with integer coefficients. Transcendental numbers are those that are not algebraic.} number $a>1$, the numbers $F(N)$ do behave as independent draws from the probability distribution $\mu$,
in the sense of averaging with the Mellin transform.   Such averaging is possible because  $F(N)$ belongs to the class of functions to which the Mellin procedure applies.

A first step to understand the statistical properties of $F(N)$  is to show that the Mellin  average of $F(N)$ vanishes.   This follows from an integration by parts argument
similar to one that was used in section~\ref{mellin}.   
We will explain a version of this argument that  applies  to all of the examples that we will
encounter.   First, without changing the output of the averaging procedure, we can use an arbitrary lower bound $N\geq N_0>0$ (rather than $N\geq 1$) in  defining the Mellin transform:
\be\label{anishing} M^{(N_0)}(s)=\int_{N_0}^\infty \d N \,N^{s-1} F(N). \ee
This does not affect the poles of the Mellin transform, which arise from the behavior for $N\to\infty$.
Now suppose that we want to average a function $e^{\i \Phi(N)}$. Suppose that $\Phi'(N)$ does not
vanish for $N>N_0$.   Using the identity $e^{\i \Phi(N)}=\frac{1}{\i \Phi'(N)}\frac{\d}{\d N}e ^{\i\Phi(N)}$ and integrating by parts,
we find 
\be\label{windfall}\int_{N_0}^\infty\d N\,N^{s-1}e^{\i \Phi(N)}=\int_{N_0}^\infty\d N \frac{\d}{\d N}\left(N^{s-1} \frac{\i}{\Phi'(N)}\right)e^{\i \Phi(N)}+\mathrm{surface~term}.\ee
The surface term at $N=N_0$ is an entire function of $s$, so it does not contribute to the poles of the Mellin transform.    There is no surface term
at $N=\infty$ if $1/\Phi'$ vanishes for $N\to\infty$ faster than any power of $1/N$.
If in addition,  $\Phi''(N)/\Phi'(N)^2$ vanishes for $N\to\infty$ 
faster than any power of $1/N$, then the integral on the right hand side of \eqref{windfall} converges for any $s$.   In this case, the  Mellin transform of
$e^{\i\Phi(N)}$ is an entire function of $s$  and the Mellin average of $e^{\i\Phi(N)}$ vanishes.   This argument applies for any $\Phi$ such that 
$1/\Phi'(N)$ and $\Phi''(N)/(\Phi'(N))^2$ both vanish at infinity faster than any power of $1/N$.   

Clearly, writing $F(N)=\cos a^N $ 
as $\tfrac{1}{2}\left( e^{\i a^N}+e^{-\i a^N}\right)$, the argument just explained applies to each of the two terms $e^{\pm \i a^N}$ and shows
that the Mellin average of $F(N)$ vanishes.
What about  higher moments of $F(N)$?   We have $F(N)^2=\tfrac{1}{2}+\tfrac{1}{2}\cos(2 a^N)$.  The averaging procedure annihilates $\cos(2 a^N)$ and leaves
the constant $\tfrac{1}{2}$ unchanged, so by our rules $\la F(N)^2\ra=\tfrac{1}{2}$.   This is the same result that we get if we treat $a^N$ as a random angle $\theta$.  The average of
$F(N)^2$ would then be
\be\label{insto} \frac{1}{2\pi}\int_0^{2\pi}\d\theta \,\cos^2\theta =\tfrac{1}{2}. \ee
More generally, for every positive integer $w$, the averaging procedure gives
\be\label{winsto}\la F(N)^w\ra =\frac{1}{2\pi}\int_0^{2\pi}\d\theta \,\cos^w\theta.\ee
To prove this, note that one way to evaluate the integral $\frac{1}{2\pi}\int_0^{2\pi}\d\theta \,\cos^w\theta$ is to make a Fourier expansion  $\cos^w\theta=\sum_r a_r e^{\i r\theta}$; then
integrating over $\theta$ discards the terms with $r\not=0$, leading to  $\frac{1}{2\pi}\int_0^{2\pi}\d\theta \,\cos^w\theta=a_0$.   For $F(N)=\cos a^N$, we make a similar expansion
$F(N)^w=\sum_r a_r e^{\i r a^N}$, with the same coefficients $a_r$.   Then for reasons just explained, the Mellin averaging
procedure discards the contributions with $r\not=0$ and sets $\la F(N)^w\ra=a_0$.   In other words, all of the moments of $F(N)$ are the same as the corresponding moments of the
function $\cos\theta$ on the unit circle.   Thus, with our averaging procedure, $F(N)$ is equivalent to the cosine of a random angle and behaves as a succession of draws from the probability distribution $\mu=\frac{\d x}{\pi\sqrt{1-x^2}}$ on the interval $[-1,1]$.

Up to this point, we could say the same for the functions $\cos (aN) $ or $\cos( a N^\alpha)$.  What distinguishes the choice $F(N)=\cos(a^N)$ is that, if $a$ is a transcendental number, then
$F(N)$ behaves as a sequence of
 {\it independent} draws from the distribution $\mu$.   For example, let us show that for any integer $p$, the values $F(N), F(N+1), F(N+2),\dots, F(N+p)$ are statistically independent,
according to our averaging procedure.    For this, we consider a joint moment $\la F(N)^{m_0} F(N+1)^{m_1}\cdots F(N+p)^{m_p}\ra$, with nonnegative integers $m_0,\dots, m_p$.
Statistical independence of the $F(N)$ means that
\be\label{statind}\la F(N)^{m_0} F(N+1)^{m_1}\cdots F(N+p)^{m_p}\ra =\prod_{r=0}^p \la F(N+r)^{m_r}\ra. \ee
To prove this, we expand
\begin{align}\label{expfun} F(N)^{m_0} F(N+1)^{m_1}\cdots F(N+p)^{m_p}&=\sum_{h_0,h_1,\dots, h_p} \alpha_{h_0 h_1\cdots h_p} \exp\left(\i\sum_{r=0}^p h_r a^{N+r}  \right)\\ &
= \sum_{h_0,h_1,\dots, h_p} \alpha_{h_0 h_1\cdots h_p} \exp\left(\i a^N \sum_{r=0}^p h_r a^{r}  \right).\end{align}
Here  $h_r$ takes values $-m_r,-m_r+2,\dots, m_r$, and $\sum_{r=0}^p h_r a^r$ is a constant independent of $N$; if $a$ is transcendental, then this constant is nonzero unless the integers $h_r$ all vanish.   For transcendental $a$, the averaging procedure will therefore 
annihilate all contributions on the right hand side of \eqref{expfun} except the one with $h_0=\cdots = h_p=0$; the average is therefore simply the coefficient $\alpha_{0,0,\dots,0}$.   
But this is exactly the behavior that we need to establish the criterion \eqref{statind}
for statistical independence of the successive values of $F$.   

It immediately follows, for example, that for any fixed integers $0\leq r_1< r_2< \cdots < r_t$, the variables $F(N+{r_1}), F(N+r_2),\dots, F(N+r_t)$ are statistically independent,
since these variables are a subset of the variables $F(N), F(N+1), \dots, F(N+r_t)$, which we already know to be statistically independent.
However, we would like to establish that, for example, $F(N)$ and $F(2N)$ or $F(N)$ and $F(N^2)$ are independent.   Such statements  can be proved by a similar argument; the main
difference is that it is not necessary to assume that $a$ is transcendental.
In general, suppose we are given increasing functions $g_0(N),\dots, g_p(N)$ with $g_0(N)/\log N\to\infty$ and with the differences $g_r(N)-g_{r-1}(N)$ being positive and diverging for 
$N\to\infty$.      To analyze the joint moment
\be\label{delf} \bigl\la F(g_0(N))^{m_0} F(g_1(N))^{m_1}\cdots F(g_p(N))^{m_p}\bigr\ra,\ee  we expand in a Fourier series 
\begin{align}\label{elf}F(g_0(N))^{m_0} &F(g_1(N))^{m_1}\cdots F(g_p(N))^{m_p}\cr&=\sum_{h_0,h_1,\dots, h_p} \alpha_{h_0,h_1,\dots,h_p} \exp\left(\i \sum_r h_r a^{g_r(N)}\right).\end{align}
For $N\to\infty$, the phase $\Phi=\sum_r h_r a^{g_r(N)}$ is dominated by a leading term $h_r a^{g_r(N)}$, with the largest value of $r$ such that $h_r\not=0$ (here
$h_r a^{g_r(N)}$ is, for large $N$, much bigger than other contributions to $\Phi$ because of the assumption that  $g_r(N)-g_{r-1}(N)$ diverges for $N\to\infty$;
 if this is not assumed, then we can proceed using the argument of  \eqref{expfun} and assuming that $a$ is
transcendental).   In particular, there is some $N_0$ such that $\Phi'\not=0$
for $N>N_0$.   Then we can define the Mellin transform by an integral on the half-line  $N\geq N_0$.   On that half-line, $1/\Phi'$ and $\Phi''/(\Phi')^2$ both vanish faster than any power of $1/N$ for $N\to\infty$.
So averaging based on the Mellin transform gives the result $ \la F(g_0(N))^{m_0} F(g_1(N))^{m_1}\cdots F(g_p(N))^{m_p}\ra=\alpha_{0,0,\dots,0}$.
That is precisely the condition that ensures that joint moments defined with the 
Mellin procedure factorize in a way that corresponds to statistical independence of $F(g_0(N)),F(g_1(N)),\dots , F(g_p(N))$.   

So we have found a simple deterministic function $F(N)=\cos(a^N)$ that simulates a sequence of independent random draws from a certain probability distribution.  This function 
oscillates in a way that is superpolynomially  rapid; that is, for large $N$, it oscillates on a scale smaller than any power of $1/N$.
  A somewhat similar function $\cos (a N^\alpha)$ that oscillates on a scale that is only polynomially small does not
achieve the same statistical independence.   Why did this happen?   We do not have a rigorous proof that nonpolynomially rapid oscillations are needed for statistical independence.
  However, a heuristic explanation is the following.   Consider a large number
$N$ and suppose that we want the values $F(N), F(N+1),\dots, F(2N)$ to be statistically independent.   
This requires the joint moments of these $N+1$ variables to factorize, which is a separate condition for each subset of them. This exponentially large number of conditions suggests that they cannot
be satisfied by a simple ansatz in the interval $[N,2N]$ involving, for example, a Fourier series with only polynomially many terms, and thus only polynomially rapid oscillations.
To make such an argument rigorous, among other things, one would have to take into account the fact that our notion of averaging and 
statistical independence is only defined asymptotically for $N\to\infty$ and
says nothing about what happens on a finite interval $[N,2N]$.   

It is interesting to compare what we have said about the function $F(N)=\cos(a^N)$ to what can be said from the standpoint of a more conventional form of averaging.
The usual procedure would be to define the average of, for example,  $ F(N)^{m_0} F(N+1)^{m_1}\cdots F(N+p)^{m_p}$ as 
\be\label{maybe}\lim_{T\to\infty}\frac{1}{T}\sum_{N=1}^T  F(N)^{m_0} F(N+1)^{m_1}\cdots F(N+p)^{m_p},\ee  if this limit exists.   Classical equidistribution results~\cite{koksma,Weyl}
imply that for {\it generic} transcendental numbers  $a>1$, these limits exist and agree with the averages defined by the Mellin procedure.  There are at least two important differences.   First,
with this classical averaging procedure, though one knows what happens for a generic transcendental number  $a$, it is very difficult to determine whether the expected statistical properties
hold for any specific case,
such as $a=\pi$ or $a=e$.   Such a question can quickly lead to unsolved problems in number theory.
 By contrast,  with the Mellin procedure, it was straightforward to establish the
statistical behavior of the function $\cos(a^N)$ for any transcendental  $a>1$.   Second, the standard averaging procedure lets one define the average of a quasirandom function
$F(N)$ or $F(N)^m$, but it does not assign a meaning to higher order terms in a $1/N$ expansion.  The Mellin procedure does assign meaning to such higher order terms.  It is true
that for averages of  the function $F(N)=\cos(a^N)$, and its powers, the Mellin procedure sets all the higher order terms in $1/N$ to zero.   This would not be true for a similar
function such as $\widetilde F(N)=\left(1+\frac{1}{N}\right) \cos(a^N)$.   The Mellin procedure will say, for example, that $\la \widetilde F(N)^2\ra =\tfrac{1}{2}\left(1+\frac{1}{N}\right)^2$.   As explained in section~\ref{mellin}, because of fluctuations, classical averaging will compute the leading term $\tfrac{1}{2}$ but will fail to assign meaning to the $1/N$ corrections.

Now we would like to modify this construction to find a deterministic function $F(N)$ whose values at integers behave like a sequence of independent draws from a Gaussian distribution
on the whole real line, rather than the distribution $\mu=\frac{\d x}{\pi\sqrt{1-x^2}}$ on the interval $[-1,1]$.   A naive approach is the following.   Let $G$ be a one-to-one mapping from the open
interval $(-1,1)$ to the whole real line which maps the distribution $\mu$ on the interval to a Gaussian distribution on the real line.    Then the values of $G(\cos(a^N))$ for integer $N$ will behave
as independent draws from a Gaussian distribution.\footnote{One should pick $a$ so that $a^N$, for integer $N$, is never an integer multiple of $\pi$; otherwise, for such values of $N$,
$G(\cos(a^N))$ is not defined.}   Unfortunately, with this simple definition, averaging based on the Mellin transform is not available.   The reason is that inevitably $G$ maps the endpoints $\pm 1$ of
the interval $[-1,1]$ to $\pm\infty$.    As a result, $G(\cos(a^N))$ has a singularity at any value of $N$ such that $a^N$ is an integer multiple of $\pi$.  In particular, these singularities are exponentially close together as  $N$ becomes large. It may be possible to define the Mellin transform of $G(\cos(a^N))$ for sufficiently negative ${\rm Re}
\,s$, but it is quite unlikely that this Mellin transform has the kind of analytic continuation that is needed for Mellin averaging. Certainly integration by parts, which is our basic tool to analyze the Mellin transform, will not give a useful result.  And intuitively, if black hole physics has a physically meaningful analytic continuation in $N$, we do not expect this continuation to produce a proliferation of singularities on the real $N$ axis.   For example, the toy models described in the
next section certainly vary smoothly with $N$.

As an alternative,  inspired by the central limit theorem, we will try to
 build a Gaussian random variable as a linear combination of many independent random variables drawn from the distribution $\mu$.  A naive try is to pick a sequence of numbers $a_1,a_2,\dots, a_N$ all greater than 1 and to define
\be\label{firsttry} F(N)=\sqrt{\frac{2}{ N}}\sum_{k=1}^N \cos(a_k^N).  \ee
For large $N$, $F(N)$ is approximately a Gaussian random variable of mean zero and variance 1.   However, the formula in \eqref{firsttry} defines $F(N)$ for a particular value of $N$;
it does not define an analytic function of $N$.   To make an analytic function of $N$ based on the same idea, we can let the sum over $k$ run over all positive integers, but include a cutoff
function in the sum so that only of order $N$ values of $k$ make important contributions to the sum.   One way to do this is to define
\be\label{secondtry} F(N)=\left({2}({e^{2/N}-1)}\right)^{1/2} \sum_{k=1}^\infty e^{-k/N} \cos(a_k^N).  \ee
(the factor $\left(2(e^{2/N}-1)\right)^{1/2}$ will ensure that $F(N)$ has variance 1).
However, it turns out that it is important to have rapid oscillations in $k$ as well as in $N$.
To achieve this we define
\be\label{finally} F(N)=\left(2(e^{2/N}-1)\right)^{1/2} \sum_{k=1}^\infty e^{-k/N} \cos(a^{ N}b^k), \ee  where the number $b>1$ will be constrained in a way that is explained shortly.   This particular choice will greatly simplify the analysis of the sum over $k$.
To define the Mellin transforms of functions of $F(N)$, we will need a power law upper bound on $|F(N)|$.   To get such a bound, we simply replace $\cos(a^{N}b^k)$ by 1, giving
\be\label{zinally}|F(N)|\leq \left(2(e^{2/N}-1)\right)^{1/2} \sum_{k=1}^\infty e^{-k/N}=\left(2(e^{2/N}-1)\right)^{1/2}\frac{1}{e^{1/N}-1}  \sim 2 N^{1/2}.\ee

As a preliminary, let us discuss why the Mellin averaging gives $\la F(N)\ra=0$.   This would be a consequence of what we have already done if one restricts the
sum over $k$ in \eqref{finally} to a finite range, but the infinite sum over $k$ raises new issues.   We want to prove that the Mellin transform
\begin{align}\label{inally} M(s)=&\int_1^\infty \d N\,N^{s-1} F(N)\cr=&\int_1^\infty \d N \,N^{s-1}  \left(2(e^{2/N}-1)\right)^{1/2}     \sum_{k=1}^\infty e^{-k/N} \cos(a^{  N}b^k)\end{align} is an entire function of $s$.
With this aim, we would like to exchange the integration over $N$ with the sum over $k$ and write
\be\label{tinally}M(s)=\sum_{k=1}^\infty \int_1^\infty \d N \,N^{s-1} \left(2(e^{2/N}-1)\right)^{1/2} e^{-k/N} \cos(a^{ N}b^k).\ee
This interchange is valid if the sum and integral are absolutely convergent.   
Because of the bound \eqref{zinally}, the sum and integral are absolutely convergent if\footnote{Similarly, in a more general correlation function such as $\la F(N+r_1) F(N+r_2) \cdots F(N+r_p) \ra$
among $p$ values of $F(N)$, the same bound shows absolute convergence of the sum and integral if ${\mathrm{Re}}\,s<-p/2$.   Note that this argument would fail if instead of building $F(N)$ as the
sum of roughly $N$ important terms, we build it in a similar fashion as the sum of, say, $2^N$ terms.   In that case the upper bound on $|F(N)|$ would
be $2^{N/2}$ instead of $N^{1/2}$, and there would be no immediately obvious region of holomorphy of the Mellin transform $M(s)$.}
${\mathrm{Re}}\,s<-1/2$.  So we will begin in the region
${\mathrm{Re}}\,s<-1/2$, where holomorphy in $s$ is manifest,
and prove an analytic continuation.   

Now that we can integrate over $N$ before the sum over $k$, we can proceed in a familiar fashion.   We write $\cos(a^{N}b^k)=\tfrac{1}{2}\left(
\exp(\i a^{N}b^k)+\exp(-\i a^{N}b^k)\right)$ and consider separately either of the two terms, say the first one.    Then inserting the identity
 $\exp(\i a^{N}b^k)=\frac{1}{\i (\log a )a^{N}b^k}\frac{\d}{\d N}
\exp(\i a^{N}b^k)$ and integrating by parts in the familiar way, we get 
\begin{align}\label{nally}M(s) =&\frac{i}{2\log a}\sum_{k=1}^\infty b^{-k} \int_1^\infty \d N   \frac{\d}{\d N}\left( \left(2(e^{2/N}-1)\right)^{1/2}   N^{s-1} e^{-k/N} a^{-N}\right)  \exp(\i a^{N}b^k)    \\ &  + {\mathrm{surface~term}}+{\rm complex~conjugate}. \end{align}
The surface term at $N=1$ is harmless as usual, because it is independent of $s$. 
Because of the factor $a^{-N}$, the  integral over $N$ is highly convergent for any  $s$.  Therefore, for fixed $k$, the integral over $N$ is  an entire function of $s$.
The factor $b^{-k}$ ensures that the sum over $k$ is highly convergent and  harmless.   So finally, $M(s)$ is an entire function, and under Mellin averaging, $\la F(N)\ra=0$.

Next let us discuss  the variance $\la F(N)^2\ra$.   For this, we apply the usual Mellin procedure to
\be\label{zefro}F(N)^2= 2(e^{2/N}-1)\sum_{k_1,k_2=1}^\infty e^{-(k_1+k_2)/N} \cos( a^{N}b^{k_1})\cos(a^{N}b^{k_2}).\ee
First,  let us explicitly evaluate the terms with $k_1=k_2=k$.  The product of cosines reduces to $\cos^2(a^{N}b^k)=\tfrac{1}{2}+\tfrac{1}{2} \cos(2 a^{N}b^k)$.
The Mellin average  of the second term $\tfrac{1}{2}\cos(2 a^{N}b^k)$ is zero, by the same analysis that showed that $\la F(N)\ra=0$.   As for the constant term $\tfrac{1}{2}$, its contribution in the
sum over $k$ can be analyzed explicitly:
\be\label{simplecontrib} 2 (e^{2/N}-1)  \sum_{k=1}^\infty e^{-2k/N}\tfrac{1}{2} =1. \ee
To show that this is the only contribution to $\la F(N)^2\ra$, so that  $\la F(N)^2\ra=1$, we have to show that the terms in \eqref{zefro} with $k_1\not=k_2$ do not contribute.   For this, we expand $\cos( a^{N}b^{k_1})\cos(a^{N}b^{k_2})$
as a sum of exponentials, each of the form $\exp\left(\i (\epsilon_1 a^{N}b^{k_1}+\epsilon_2 a^{N}b^{k_2})\right)$, with $\epsilon_i=\pm 1$.   The phase is
$\Phi= \epsilon_1 a^{N}b^{k_1}+\epsilon_2 a^{N}b^{k_2}$, so
\be\label{zono}\frac{1}{\d\Phi/\d N}= a^{-N}\frac{1}{\log a}\frac{1}{ \epsilon_1 b^{ k_1} +\epsilon_2 b^{ k_2}}. \ee
After integrating by parts in the usual way in the  Mellin transform of $F(N)^2$, the factor $a^{-N}$ will ensure that the integral over $N$ converges for all $s$, so the Mellin transform is an entire
function of $s$ if the sum over the $k_i$ is convergent.
This is true because of   the factor $1/\Theta$ in $(\d\Phi/\d N)^{-1}$, with $\Theta={ \epsilon_1 b^{ k_1} +\epsilon_2 b^{k_2}}$.   First of all,  $\Theta$ never vanishes for $k_1\not= k_2$, so
each individual contribution in the sum over the $k_i$ is finite.   Now let us see what happens if either or both of the $k_i$ is large.
   Without essential loss of generality, we can assume that $k_1<k_2$, and we set $k_1=k_{\min}$, $k_2-k_1=\Delta k$.  Then
\be\label{ziflo}\Theta= {b^{ k_{\min}}}({\epsilon_1 +\epsilon_2 b^{ \Delta_k}}). \ee
Clearly, $\Theta$ grows exponentially if either $k_{\min}$ or $\Delta k$ becomes large, and therefore the sum over the $k_i$ converges exponentially fast.   Hence the contributions to $F(N)^2$ with
$k_1\not=k_2$ do not contribute to the Mellin average $\la F(N)^2\ra$.

Now let us analyze a general average\footnote{We could set one of the $r_i$ to vanish without essential loss of generality; to keep the formulas symmetrical, we do not do that.}
     $\la F(N+r_1) F(N+r_2) \cdots F(N+r_p) \ra$  with fixed integers $r_i$, not necessarily distinct.\footnote{One could replace $N+r_i$ by general increasing  functions $g_i(N)$ where some or all of the differences $g_i(N)-g_{i-1}(N)$ diverge as $N\to\infty$ while the others remain fixed.  This case can be treated similarly, along lines
explained earlier in the analysis of the function $\cos(a^N)$.}  Expanding the product of cosines $\prod_{i=1}^p\cos(a^{N+r_i}b^{k_i})$ as a sum of exponentials, we encounter the phase 
\be\label{phasefactor} \Phi=a^N\left(\epsilon_1 b^{k_1} a^{r_1}+\epsilon_2 b^{k_2}a^{r_2}+\cdots +\epsilon_p b^{k_p}a^{r_p}\right),\ee
with $\epsilon_i=\pm 1$.   We do not assume the $k_i$ to be distinct, just as we have not assumed the $r_i$ to be distinct.   Exceptionally, if the $k_i$ and $r_i$ are equal in pairs while the
corresponding $\epsilon_i$ are equal and opposite, there is a complete cancellation, and $\Phi=0$, independent of $a$ and $b$.   If $a$ and $b$ are chosen generically, then
$\Phi$ never vanishes except in this case of complete cancellation.   Indeed, assuming that $\Phi$ is not identically zero, the condition $\Phi=0$ is a polynomial equation in $a$ and $b$ with integer
coefficients.   The overall factor $a^N$ in $\Phi$ does not affect this equation, which depends only
on the other data.  (A common integer shift in all $k_i$ is similarly irrelevant,)
    As we vary the positive integers $p$, $r_i$, and $k_i$ and the signs $\epsilon_i$, all polynomials in $a$ and $b$ with integer coefficients arise.    The pair of numbers $a,b$ is said to be algebraically independent over
    $\mathbb{Q}$ if none of these polynomials vanish.   This concept generalizes the notion of a transcendental number.
    Clearly, if $a$ and $b$ are algebraically independent over $\mathbb{Q}$, then by considering polynomials that depend only on $a$ or only on $b$, we see in particular that $a$ and $b$ are transcendental.
    A generic pair of numbers is algebraically independent over $\mathbb{Q}$, just as a generic number is transcendental.\footnote{There are only countably many polynomial equations with integer coefficients in a given number of variables.   So there are only countably many numbers $a$
    that satisfy one of those equations, and therefore a generic number $a$ satisfies none of them and is transcendental.   Similarly for a given transcendental number $a$, there are only countably many $b$ such that the pair $a,b$ is not algebraically independent over $\mathbb{Q}$, and therefore a generic pair is independent.}   However, it is notoriously hard to decide if a specific pair of numbers such as $a=e$, $b=\pi$ is algebraically independent over $\mathbb{Q}$.

Continuing the analysis of \eqref{phasefactor}, we will show that
if  $a,b$ are algebraically independent over $\mathbb{Q}$, then the poles of the Mellin transform
of  $F(N+r_1) F(N+r_2) \cdots F(N+r_p)$ come entirely from contributions in which $\Phi$ is identically zero.    Let us first restrict to values of $k_i$, $\epsilon_j$ for
which $\Phi$ does not vanish identically.  (Unless the $r_i$ are equal in pairs, this restriction is actually vacuous.)  
In general, some of the $k_i$ may be equal, so  the $p$ numbers $k_1,k_2,\dots, k_p$ may take only $p'<p$ distinct values.   If so, we can reduce to the case that the $k$'s are all distinct by combining together terms
with the same value of $k$.  If the combined coefficient of some $b^k$ vanishes, we drop that $k$. The sum over such $k$ converges because of the factors $e^{-k/N}$ in $F(N)$, giving at most a power of $N$, which is overcome by the factor $a^{-N}$ below. Then we have
\be\label{factor2}\Phi=a^N\Theta,~~~~ \Theta= \left(w_1 b^{ k_1}+w_2 b^{k_2} +\cdots +w_{p'} b^{k_{p'}}\right), \ee
where $w_1,\dots, w_{p'}$ are obtained by combining terms $\epsilon_i a^{r_i}$ with the same value of $k_i$. (Thus, the $w_i$ are polynomials in $a$ with integer coefficients and in particular are nonzero.)  So
\be\label{actor2}\frac{1}{\Phi'}=a^{-N}\frac{1}{\log a} \frac{1}{\Theta}.\ee   Algebraic independence over $\mathbb{Q}$ of the pair $a,b$ ensures that  $\Theta\not=0$ 
and given this,  the factor $a^{-N}$ ensures that for fixed $k_i$,  the integral over $N$ defines an entire function of $s$.   If, therefore, the sum over the $k_i$ is sufficiently convergent,
contributions in which $\Phi$ is not identically zero do not contribute to the Mellin average.

We will show that the sum over the $k_i$ is highly convergent by showing that $\Theta$ diverges exponentially as any of the $k_i$ go to infinity.   This follows by an induction on $p'$.
For $p'=1$, the statement is trivial: in that case $\Theta = w b^{ k}$ contains only one term 
and certainly diverges for $k\to\infty$.    For $p'>1$, since we have reduced to the case that the $k_i$ are distinct,
one of them is the smallest, and without essential loss of generality, we can assume that this is $k_{p'}=k_{\min}$.   Then we define $\Delta k_i =k_i-k_{\min}$ for $i=1,\dots, p'-1$.  So 
$k_{\min}$ and $\Delta k_i$ are all strictly positive and  $\Theta= b^{ k_{\min}}\Theta'$ with
\be\label{ctor2} \Theta' =\left( w_1 b^{ \Delta k_1}+w_2 b^{ \Delta k_2}+\cdots + w_{p'-1} b^{ \Delta k_{p'-1}} +w_{p'}\right). \ee
As some or all of the $\Delta k_i$ becomes large, some terms in $\Theta'$ grow exponentially and the remaining terms (namely $w_{p'}$, and any terms $w_t b^{\Delta k_t}$ where
$\Delta k_t$ is not becoming large) become negligible in comparison.   
Thus as some of the $\Delta k_i$ become large, $\Theta'$ reduces to a function of the same form as  the original function $\Theta$ defined in equation \eqref{factor2}
but with $p'$ replaced by a smaller positive integer.   By the induction hypothesis, $\Theta'$ grows exponentially in any such limit.   Since $\Theta'$ never vanishes  and
becomes large as any of the $\Delta k_i$ becomes large, its absolute value has a positive lower bound $|\Theta'|\geq \delta>0$.   So $|\Theta|\geq b^{k_{\min}}\delta$ grows exponentially
as $k_{\min}\to\infty$, and of course it also grows exponentially when $\Theta'$ grows exponentially.   Altogether, we see that $\Theta$ grows exponentially as any of the $k_i$ become large, as claimed.

So the Mellin average $\la F(N+r_1) F(N+r_2) \cdots F(N+r_p) \ra$ receives contributions only from values of the $k_i, \epsilon_j$ such that the phase $\Theta$ vanishes identically.
Such contributions only exist if the $r_i$ are equal in pairs.    So in fact, the nonzero Mellin averages are of the form \be\label{helpful}\la F(N+r_1)^{m_1}  F(N+r_2)^{m_2}\cdots F(N+r_p)^{m_p}\ra,\ee
where now we assume that the $r_i$ are all distinct, and the $m_i$ must all be even to get a nonzero average.   We want to establish that different values of $F$ satisfy the familiar
criterion of statistical independence:
\be\label{tor2}\bigl\la F(N+r_1)^{m_1}  F(N+r_2)^{m_2}\cdots F(N+r_p)^{m_p}\bigr\ra =\prod_{t=1}^p \la F(N+r_t)^{m_t}\ra. \ee
This is actually an immediate consequence of the fact that contributions to the Mellin average come only from choices of $k_i,\epsilon_j$ such that  $\Theta$ vanishes identically.
Using the definition \eqref{finally} of $F(N)$, we  expand each factor  $F(N+r_t)^{m_t}$ as a sum over $m_t$ pairs $k^{(t)}_j, \epsilon^{(t)}_j$.
The condition for $\Theta$ to vanish identically places a very strong condition on the $k^{(t)}_j, \epsilon^{(t)}_j$ for each $t$, namely the $k^{(t)}_j$ must be equal in pairs with
the corresponding $\epsilon^{(t)}_j$ being equal and opposite.   But notably, this condition establishes no correlation between the allowed choices of $k^{(t)}_j,\epsilon^{(t)}_j$ with
different values of $t$, and each set of allowed choices just makes a factorized contribution to the Mellin average.    The factorization \eqref{tor2} is immediate.

Thus, the $F(N)$ with different values of $N$ behave in the Mellin sense as statistically independent  variables.   But they are not quite {\it Gaussian} variables; they become Gaussian
only in the large $N$ limit.  To see this, let us compare the moments $\la F(N)^2\ra$ and $\la F(N)^4\ra$.   If $x$ is a Gaussian random variable with mean zero, then $\la x^4\ra=3\la x^2\ra^2$.
So a claim that $F(N)$ is a Gaussian random variable would assert that
\be\label{predgauss} \la F(N)^4\ra \overset{?}{=} 3\la F(N)^2\ra^2. \ee
Is this so?  In the formula \eqref{zefro} for $F(N)^2$, the terms that contribute to $\la F(N)^2\ra$ are the terms with $k_1=k_2=k$.   The contribution of such a term to
$\la F(N)^2\ra$ is
\be\label{diagcont} 2(e^{2/N}-1)\cdot e^{-2k/N}\tfrac{1}{2}.\ee
Summing this over $k$ gives 1.     We can, of course, write a similar formula for $F(N)^4:$ 
\be\label{tefro}F(N)^4= 4(e^{2/N}-1)^2\sum_{k_1,\dots, k_4=1}^\infty \exp\left(-\sum_{i=1}^4 k_i/N\right)\prod_{i=1}^4 \cos( a^{ N}b^{k_i}).\ee
The only contributions to $\la F(N)^4\ra$ will come from the case that the $k_i$ are equal in pairs, say $k_1=k_2=k$, $k_3=k_4=k'$, or permutations of such an arrangement.
Including a factor of 3 because $k_1$ could have been paired with $k_3$ or $k_4$ instead of $k_2$, the contribution of such a set of $k_i$, {\it as long as $k\not=k'$,} is
\be\label{efro} 3\cdot 4 (e^{2/N}-1)^2 \cdot e^{-2k/N}\tfrac{1}{2}\cdot e^{-2k'/N}\tfrac{1}{2}.\ee
So far, this is consistent with $F(N)$ being precisely a Gaussian variable.  But the exceptional case $k=k'$ produces a discrepancy.   The Gaussian prediction for the contribution
from $k=k'$ is
\be\label{fro} 3\cdot 4 (e^{2/N}-1)^2 \cdot e^{-4k/N}\frac{1}{4}.\ee
But, taking into account the fact that $\la \cos^4\theta\ra=\frac{3}{8}$, the contribution to $\la F(N)^4\ra$ from $k=k'$ is actually 
\be\label{tro} 4 (e^{2/N}-1)^2 \cdot e^{-4k/N}\frac{3}{8}.\ee
The difference is $-4(e^{2/N}-1)^2 \cdot e^{-4k/N}\frac{3}{8}$.  Summing this over $k$, we learn that
\begin{align}\label{discrep} \la F(N)^4\ra -3\la F(N)^2\ra^2&=-4(e^{2/N}-1)^2 \sum_{k=1}^\infty e^{-4k/N}\frac{3}{8}=-\frac{3}{2}\frac{(e^{2/N}-1)^2}{e^{4/N}-1}\\ &=-\frac{3}{2N}+{\mathcal O}\left(\frac{1}{N^2}\right). \end{align}

Thus, the variables $F(N)$ behave not as independent draws from a fixed probability distribution $\nu$, but as independent draws from an $N$-dependent family of probability distributions $\nu_N$
that converge in the large $N$ limit to a Gaussian with zero mean and unit variance, but with non-Gaussian corrections  that can be described by an asymptotic series in powers of 
$1/N$. 

At first sight, this is disturbing.
  In the context of gravity, normally one does not want non-Gaussian corrections that appear as a power series in $1/N$.  Non-Gaussian corrections would come in gravity from 
multi-boundary wormholes, and one would expect them to be exponentially suppressed in $N$. 
However, a simple redefinition of the $F(N)$ converts them into Gaussian random variables.     There is a unique $N$-dependent reparametrization of the real line of the
form $G_N(x)=x+ \frac{d_1(x)}{N}+\frac{d_2(x)}{N^2} + \cdots$ that maps the $N$-dependent distribution $\nu_N$ to the fixed Gaussian distribution $\nu$.  Here $d_i(x)$ is a polynomial
in $x$ of degree $2i+1$, and moreover is odd under $x\to -x$.  The first of these polynomials is  $d_1(x)=(x^3-3x)/16$.    With the aid of the mapping $G_N(x)$, we can define
a new function $\widetilde F(N)=G_N(F(N))$ whose values at integer values of $N$ behave as independent draws from a fixed Gaussian ensemble.

There is a key difference here from the attempt discussed earlier to map the distribution $\mu$ on the interval $[-1,1]$ to a Gaussian distribution on the real line.    When one maps an interval
to the whole real line, one inevitably produces singularities at the endpoints of the interval, and these singularities would prevent us from applying the Mellin averaging
procedure.   By contrast, in the present context, we are mapping the whole real line to itself and there is no such difficulty.

An important point is that the mapping $G_N$ and indeed the family $\nu_N$ of probability distributions that we are trying to map to a fixed Gaussian distribution only exist in the sense of a 
$1/N$ expansion.   In the present problem, probabilities are purely an emergent concept for large $N$.   For finite $N$, we simply have a deterministic function $F(N)$; there are no probabilities.
Probabilities and the family $\nu_N$ of probability distributions emerge in the large $N$ limit.   There is no reason to believe that the series $G_N(x)=x+\frac{d_1(x)}{N}+\frac{d_2(x)}{N^2}+\cdots$
converges or that it actually defines an $N$-dependent family of diffeomorphisms of the real line.  However, to any order in the $1/N$ expansion, there is no problem with a change
of variables such as $x\to x+\frac{d_1(x)}{N}+\frac{d_2(x)}{N^2}+\cdots$.   Such a change of variables is invertible order by order in $1/N$ and does behave in the $1/N$ expansion
 like a diffeomorphism of the real line.

A final comment is the following.  In classical probability theory, given a sequence of random variables $F(N)$ defined only for positive integer $N$, with no additional structure, it would not make
sense to claim that $F(N)$ is drawn independently for each $N$ not from a fixed probability distribution $\nu$ but from a family of distributions $\nu_N$ that
differ from $\nu$ by an amount of order $1/N$.
 If the difference between $\nu_N$ and $\nu$ is of order $1/N$, then the relative entropy between them is of order $1/N^2$.
As $\sum_{N=1}^\infty \frac{1}{N^2}<\infty$, a classical theorem of Kakutani~\cite{Kakutani} implies that no statistical test, no matter how powerful, can determine whether the sequence
$F(N)$ was drawn from the fixed distribution $\nu$ or from the sequence $\nu_N$.   Intuitively, this is because likely fluctuations in the sequence $F(N)$, if it is drawn from the fixed distribution
$\nu$, are greater than the expected effects of drawing the $F(N)$ from the family of distributions $\nu_N$ rather than the fixed distribution $\nu$.   In the context of the present article, the two
possibilities can be distinguished, because there is much more
structure: we assume that $F(N)$ is in a class of functions to which the Mellin averaging applies.

\section{Toy Models of Analytic Continuation in N}\label{toyanalytic}

Here we will describe some toy models of analytic continuation in $N$.   We will represent a black hole by, roughly, a chaotic spin chain with $N$ spins.  
When $N$ changes, the number of spins or qubits representing the black hole will have to change.  How can this happen continuously?

Our answer to this question will be to model not a black hole in isolation, but a black hole interacting with particles and fields outside the black hole,
in a spacetime that is either asymptotically Anti de Sitter or asymptotically flat.   The total number of qubits will be infinite, because infinitely many qubits
are needed to describe particles and fields outside the black hole.   As we increase $N$, the Hilbert space will remain fixed, but the number of qubits that
participate in describing the black hole will increase.

As a first step, consider a semi-infinite chain of qubits, labeled by $i=1,2,3,\dots$.   We consider a simple Hamiltonian
\be\label{simpleh} H_0=\sum_{i=1}^\infty h_i, \ee
where $h_i$ is, for example, a Hamiltonian that acts only on qubits $i$ and $i+1$.    We take the $h_i$ to be all the same (except for the labels of the
spins they act on) and sufficiently generic so that $H_0$ describes a semi-infinite chaotic spin chain.   

We would like to modify the model so that, approximately, the chaotic spin chain only consists of the first $N$ qubits.   A crude way to do this is the following.
Let $\Pi_i$ be an operator that projects the $i^{th}$ qubit onto a pure state which we call the ground state; 
we take the same pure state for all $i$.   Then consider the Hamiltonian
\be\label{zillow} H(N)=\sum_{i=1}^\infty h_i +\Delta \sum_{i=1}^\infty e^{c(i-N)}(1- \Pi_i),   \ee
where $c>0$ is a dimensionless constant and $\Delta>0$ has dimensions of energy.   For $i\ll N$, the perturbation involving $\Pi_i$ is negligible and the $i^{th}$ qubit is part of a chaotic spin chain.  We interpret
this initial part of the chain as a model of the black hole.   For $i\gg N$, there are infinitely many qubits  that at reasonably low energy are all frozen in their ground states.
We interpret those frozen qubits as describing the world outside the black hole.   Of course, the qubits with $i\gg N$ do have excitations, with an excitation energy $\Delta e^{c(i-N)}$ that grows rapidly with $i$.
We regard this as a very crude model of the following basic fact: in Anti de Sitter space, the energy of a particle increases as it approaches the boundary, which
we model here by $i\to\infty$.  One could make the model more realistic by adding to
$1-\Pi_i$ an additional term that would enable an excitation at site $i$ to ``hop'' to a neighboring site (as in the Heisenberg ferromagnet discussed
presently).    One could also, of course, replace the function
$e^{c(i-N)}$ by some other rapidly increasing function.   

In what sense do new degrees of freedom appear as $N$ increases? Because the summand in the definition of $H(N)$ depends only on the difference
$i-N$, increasing $N$ by 1 has exactly the same effect as taking the lower bound on $i$ to be 0 instead of 1.   Therefore, up to relabeling the qubits,
increasing $N$ by 1 has no effect on the world outside the black hole, but adds one qubit to the black hole.  

An inelegant feature of the model as defined so far is that the chaotic part of the Hamiltonian $\sum_i h_i$ contributes to the Hamiltonian of the qubits at $i\gg N$,
far outside the black hole.   Though this contribution is negligible, it is not well-motivated physically.   We can eliminate this contribution by including
a smooth cutoff function such as $\tfrac{1}{2}\left(1-\tanh(i-N)\right)$, chosen to be  very nearly 1 for $i\ll N$ and $0$ for $i\gg N$.  An improved model
is thus
\be\label{nillow} H(N)=\sum_{i=1}^\infty h_i \tfrac{1}{2}\left(1-\tanh(i-N)\right)+\Delta \sum_{i=1}^\infty e^{c(i-N)}(1- \Pi_i ) . \ee

To make a similar model of a black hole in an asymptotically flat spacetime, we want the excitation energies outside the black hole to be independent
of the distance from the black hole.  For this we simply replace the cutoff function $e^{c(i-N)} $ by, for example, $\frac{1}{2}\left(1+\tanh(i-N)\right)$, to get
\be\label{willow} H(N)=\sum_{i=1}^\infty h_i \tfrac{1}{2}\left(1-\tanh(i-N)\right)+\Delta \sum_{i=1}^\infty \tfrac{1}{2}\left(1+\tanh(i-N)\right)(1- \Pi_i) .  \ee
Flipping a qubit at $i\gg N$ now involves a fixed excitation energy $\Delta$.
Again, to make this more realistic, we should allow excitations outside the black hole to hop between sites.   A simple, explicit model for this is a Heisenberg
ferromagnet in a magnetic field, with Hamiltonian  $\Delta \sum_i (1-\sigma_{i,z}) +\mu \sum_i(1- \vec \sigma_i\cdot \vec\sigma_{i+1})$, where $\vec\sigma_i$ are Pauli matrices of the $i^{th}$ spin.   For $\Delta,\mu>0$,
the ground state has
$\sigma_z=1$ for all spins; the quasiparticle excitations are magnons, involving the flip of a single spin.    A model of a black hole in an asymptotically
flat spacetime with somewhat more realistic modeling of the physics outside the black hole is thus 
\begin{align}\label{illow} H(N)=&\sum_{i=1}^\infty h_i \tfrac{1}{2}\left(1-\tanh(i-N)\right)\cr &+\sum_{i=1}^\infty 
\tfrac{1}{2}\left(1+\tanh(i-N)\right) \left(\Delta (1- \sigma_{i,z}) +\mu (1- \vec \sigma_i\cdot \vec\sigma_{i+1} ) \right)  .  \end{align}
Of course, we could also include a parameter controlling the width of the transition region from the black hole to the outside world,
analogous to the parameter $c$ in \eqref{zillow}.  We could also replace the simple spin chain that has been assumed with a $d$-dimensional arrangement of
spins, for any $d$.

In these models, the black hole entropy and energy {\it at a fixed temperature} are asymptotically linear in $N$.   The precise behavior is determined by the thermodynamic limit of the spin chain in question at the given temperature.

The Hilbert space of these models does not decompose precisely
as the tensor product of a black hole Hilbert space and a Hilbert space
describing the outside world.   That is actually a desirable feature of the models, as typically in black hole physics no such precise decomposition exists.   Supersymmetric 
or BPS black holes are the outstanding example in which one can indeed define a Hilbert space that describes  the states of the black hole and nothing else.
The ideas that we have presented here would not suffice for analytically continuing a BPS black hole as a function of $N$.

\section*{Acknowledgments}
We thank Yiming Chen, Jordan Cotler, Victor Ivo, Wei Li, Henry Lin, Hong Liu,  Vladimir Narovlansky, Steve Shenker, and Brian Swingle for discussions. We are particularly grateful to Juan Maldacena for many clarifying conversations. JKF is supported by the Marvin L. Goldberger Member Fund at the Institute for Advanced Study and the National Science Foundation under Grant PHY-2514611. Research of EW is partly supported by NSF-2514611.

\appendix 

\section{Combinatorial Approach}
\label{sec:combinatorics}

In this appendix, we provide a complementary proof  of \eqref{eq:chebconn}, using combinatorial tools. In this approach, the connection to the double cone wormhole is  more manifest.

We consider $H$ to be an $L \times L$ GUE matrix and $PHP$ is a $\alpha L \times \alpha L $ principal submatrix, for instance the upper left corner. The correlations of the matrix elements are Gaussian
\begin{align}
    \la H_{ab}\ra = 0, \quad \la H_{ab}H_{cd}\ra = \frac{1}{L}\delta_{ad}\delta_{bc},
\end{align}
so that $H$ has a semicircle distribution of radius $2$.  The division of $PHP$ by $\sqrt{\alpha}$ uniformizes the spectrum such that $\tfrac{PHP}{\sqrt{\alpha}}$ is also a semicircle distribution of radius 2. 

Using Wick's theorem, $\la \Tr(H^n)\Tr((PHP)^m)\ra$ may be evaluated as a sum over pairings, $\pi$, of $n+m$ (even) elements
\begin{align}
    \la \Tr(H^n)\Tr((PHP)^m)\ra = \sum_{\pi} L^{C(\gamma\circ \pi) -(n+m)/2}\alpha^{C_{K}(\gamma\circ \pi)},
\end{align}
where $C$ is the number of cycles in the permutation, $\gamma = (1,\dots,n)(n+1,\dots,n+m)$ in cycle notation representing the traces, and the factor of $(n+m)/2$ comes from the normalization of matrix elements. $C_K$ is the number of cycles for permutations that involve the $PHP$ trace. 
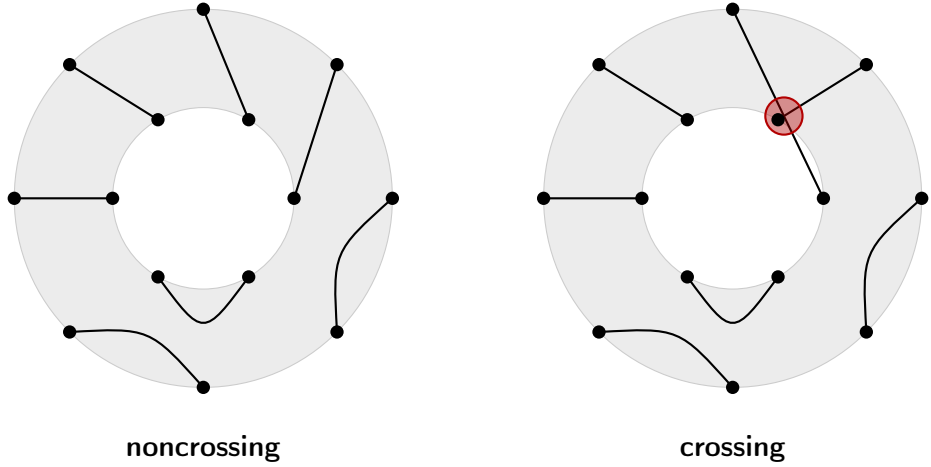
\begin{figure}
\centering
\begin{tikzpicture}
\def\Ro{2.5}    % outer radius
\def\ri{1.2}    % inner radius
\def\mon{1.85}   % control-point radius (middle of annulus)
\def\sep{7}    % horizontal gap between diagrams
% ================================================================
%  LEFT — noncrossing
% ================================================================
\begin{scope}
  % annulus
  \fill[gray!15] (0,0) circle (\Ro);
  \fill[white]   (0,0) circle (\ri);
  \draw[gray!40] (0,0) circle (\Ro);
  \draw[gray!40] (0,0) circle (\ri);
  % 8 outer points (every 45°)
  \foreach \i/\a in {1/90,2/45,3/0,4/-45,5/-90,6/-135,7/180,8/135}
    {\coordinate (LO\i) at (\a:\Ro);}
  % 6 inner points (every 60°)
  \foreach \i/\a in {1/60,2/0,3/-60,4/-120,5/180,6/120}
    {\coordinate (LI\i) at (\a:\ri);}
  % --- arcs (all black, all routed through the annulus) ---
  % outer–inner (naturally cross the annulus)
  \draw[black, thick] (LO1) -- (LI1);
  \draw[black, thick] (LO2) -- (LI2);
  \draw[black, thick] (LO7) -- (LI5);
  \draw[black, thick] (LO8) -- (LI6);
  % outer–outer (control point pulls arc inward into annulus)
  \draw[black, thick] (LO3) .. controls (-22.5:\mon) .. (LO4);
  \draw[black, thick] (LO5) .. controls (-112.5:\mon) .. (LO6);
  % inner–inner (control point pushes arc outward into annulus)
  \draw[black, thick] (LI3) .. controls (-90:\mon) .. (LI4);
  % dots (on top)
  \foreach \i in {1,...,8}{\fill (LO\i) circle (2.5pt);}
  \foreach \i in {1,...,6}{\fill (LI\i) circle (2.5pt);}
  \node[font=\sffamily\bfseries, anchor=north] at (0,-\Ro-0.5)
    {noncrossing};
\end{scope}
% ================================================================
%  RIGHT — crossing
% ================================================================
\begin{scope}[xshift=\sep cm]
  % annulus
  \fill[gray!15] (0,0) circle (\Ro);
  \fill[white]   (0,0) circle (\ri);
  \draw[gray!40] (0,0) circle (\Ro);
  \draw[gray!40] (0,0) circle (\ri);
  % same point positions
  \foreach \i/\a in {1/90,2/45,3/0,4/-45,5/-90,6/-135,7/180,8/135}
    {\coordinate (RO\i) at (\a:\Ro);}
  \foreach \i/\a in {1/60,2/0,3/-60,4/-120,5/180,6/120}
    {\coordinate (RI\i) at (\a:\ri);}
  % --- arcs (same pairing EXCEPT O1,O2 partners are swapped) ---
  % the two crossing chords
  \draw[black, thick, name path=c1] (RO1) -- (RI2);
  \draw[black, thick, name path=c2] (RO2) -- (RI1);
  % everything else identical to the left diagram
  \draw[black, thick] (RO3) .. controls (-22.5:\mon) .. (RO4);
  \draw[black, thick] (RO5) .. controls (-112.5:\mon) .. (RO6);
  \draw[black, thick] (RO7) -- (RI5);
  \draw[black, thick] (RO8) -- (RI6);
  \draw[black, thick] (RI3) .. controls (-90:\mon) .. (RI4);
  % mark the crossing
  \path[name intersections={of=c1 and c2, by=X}];
  \fill[red!70!black, opacity=0.4] (X) circle (7pt);
  \draw[red!70!black, thick] (X) circle (7pt);
  % dots (on top)
  \foreach \i in {1,...,8}{\fill (RO\i) circle (2.5pt);}
  \foreach \i in {1,...,6}{\fill (RI\i) circle (2.5pt);}
  \node[font=\sffamily\bfseries, anchor=north] at (0,-\Ro-0.5)
    {crossing};
\end{scope}
\end{tikzpicture}
    \caption{Examples of two annular pairings with $n = 8$ and $m = 6$. The left pairing is noncrossing while the right is not.}
    \label{fig:ANC}
\end{figure}

Only the pairings that connect the two traces contribute to the connected correlator. $C(\gamma\circ \pi)$ is maximized by annular noncrossing partitions connecting the two traces, $\mathrm{ANC}_c$, which have $C(\gamma\circ \pi) =(n+m)/2$. Annular noncrossing pairings are those that are noncrossing when drawing one trace as a circle inside of another circle representing the other trace (defining the annulus or cylinder). An example is shown in figure~\ref{fig:ANC}. Therefore, for large $L$, we have
\begin{align}
    \la \Tr(H^n)\Tr((PHP)^m)\ra_c = \sum_{\pi \in \mathrm{ANC}_c} \alpha^{C_{K}(\gamma\circ \pi)} +O(L^{-2}).
\end{align}
We will drop the $O(L^{-2})$ subsequently.

In a given connected pairing, there are $\min(n,m)\geq r(\pi) \geq 1$ chords connecting the $H$ trace to the $PHP$ trace. Moreover, both $n-r$ and $m-r$ must be even so that there is a noncrossing pairing in the remaining trace. Thus, $n$, $m$, and $r$ must have the same even-odd parity. Any of the $r$ through connections contribute to $C_{K}(\gamma\circ \pi)$ as well as the $(m-r)/2$ pairings of the $PHP$ trace\footnote{This is the one place where this calculation differs from two correlated GUE matrices of the same size, with $\la (H)_{ab}(PHP)_{cd}\ra = \frac{\alpha}{L}\delta_{ad}\delta_{bc}$. There, no $\alpha$ contribution would come from the $(m-r)/2$ pairings of the $PHP$ trace.}, so 
\begin{align}
    \la \Tr(H^n)\Tr((PHP)^m)\ra_c = \sum_{\pi \in \mathrm{ANC}_c} \alpha^{(m+r(\pi))/2}.
\end{align}

We are thus tasked with counting the number of annular noncrossing partitions with $n$ and $m$ insertions on the two circles and $r$ chords spanning across.
The $r$ chords separate the $H$ trace into sections of $2p_1,\dots,2p_r$ $H$ insertions with $2\sum_{i}p_i = n-r$ and similarly for $PHP$, it is separated into sections of $2q_1,\dots,2q_r$ $PHP$ insertions with $2\sum_i q_i = m-r$. The number of noncrossing pairings of each section is given by the Catalan number
\begin{align}
    C_{p} = \frac{1}{p + 1}\binom{2p}{p}.
\end{align}
The number of noncrossing pairings of $n$ insertions with $r$ insertions previously distinguished is
\begin{align}
    c_{n,r} = \sum_{2\sum_{i}p_i = n-r}\prod_{i=1}^rC_{p_i},
\end{align}
where the sum is over all ways for $2\sum_{i}p_i$ to sum to $n-r$. We have chosen a particular point as the root, so we should multiply by $n$. This furthermore overcounts by a factor of $r$ because the root chord could be any of the $r$, so we will be interested in $\tfrac{n}{r}c_{n,r}$. To evaluate this sum, we first consider the Catalan generating function
\begin{align}
    C_a(z) = \sum_{p =0}^{\infty} C_p z^p=\frac{1-\sqrt{1-4z}}{2z}
\end{align}
and take it to the $r^{th}$ power
\begin{align}
    C_a(z)^{r} = \sum_{p_1,\dots , p_r =0}^{\infty} z^{\sum_i p_i} \prod_{i=1}^r C_{p_i}
\end{align}
$c_{n,r}$ is then the $\tfrac{n-r}{2}$ coefficient in the power series, which is
\begin{align}
     c_{n,r} = \frac{r}{n}\binom{n}{\frac{n-r}{2}}.
\end{align}
Then, including the cyclic permutation of the $r$ chords, which counts the equivalent ways to connect the two traces (the analog of the zero mode of the double cone that gives the factor of $T$), the total number of $\mathrm{ANC}_c$ with $r$ chords is 
\begin{align}
    N_{n,m}(r)= r\frac{n m }{r^2} c_{n,r}c_{m,r} = r\binom{n}{\frac{n-r}{2}}\binom{m}{\frac{m-r}{2}}.
\end{align}
Thus,
\begin{align}
    \la \Tr(H^n)\Tr((PHP)^m)\ra_c = \sum_{1\leq r\leq \min(n,m)}r\binom{n}{\frac{n-r}{2}}\binom{m}{\frac{m-r}{2}}\alpha^{(m+r)/2}
\end{align}
where the sum is only over cases with $n$, $m$, and $r$ of the same parity.

As in the main text, it is useful to use Chebyshev polynomials of the first kind, which are defined by
\begin{align}
    T_r(\cos \theta) = \cos(r\theta).
\end{align}
From these, we define the matrix polynomials
\begin{align}
    X_r \equiv \Tr(T_r\left(\frac{H}{2}\right)), \quad Y_r \equiv \Tr(T_r\left(\frac{PHP}{2\sqrt{\alpha}}\right)) ,
\end{align}
where the factor of $2$ (or $2\sqrt{\alpha}$) is because the polynomial is defined on the interval $[-1,1]$.
The key point is that monomials have coefficients $d_{n,r} \equiv \binom{n}{\frac{n-r}{2}}$ when expanded in terms of the Chebyshev polynomials
\begin{align}
    x^n =
        d_{n,0}+ 2\sum_{1\leq r \leq n}d_{n,r}T_r\left(\frac{x}{2}\right)
\end{align}
and thus
\begin{align}
    4\sum_{1\leq p,q\leq n,m}d_{n,p}d_{m,q}\alpha^{m/2}\la X_p Y_q\ra_c = \sum_r  rd_{n,r}d_{m,r}\alpha^{(m+r)/2},
\end{align}
where the constant term has been dropped because it has no connected piece.
From here, we read off 
\begin{align}
    \la X_rY_s\ra_c = \frac{\delta_{rs}r \alpha^{r/2}}{4} + O(L^{-2}),
\end{align}
which matches \eqref{eq:chebconn}.

\section{Quartic Potential}
\label{sec:quartic}
In section~\ref{sec:loop}, we discussed the GUE even though \eqref{eq:Wa} was valid for general one-cut matrix models. Here, we work out the quartic potential $V(x) = \tfrac{1}{2}x^2 + \frac{g}{4}x^4$ as a more nontrivial consistency check.
The general one-cut loop equation gives
\begin{align}
    G_H(x)^2 - V'(x) G_H(x)+P(x) = 0
\end{align}
where $P(x)$ is a degree $\mathrm{deg}(V) -2$ polynomial that is fixed by $G_H(x) \rightarrow 1/x$ at large $x$. The general solution can be written as
\begin{align}
    G_H(x) = \tfrac{1}{2}\left(V'(x) -M(x) \sqrt{(x-a)(x-b)} \right)
\end{align}
where $M(x) $ is a degree $\mathrm{deg}(V) -2$ polynomial. Because the quartic is symmetric, we can take $b = -a  = 2d$ and parametrize $M(x) = m_2 x^2 + m_0$, so
\begin{align}
    G_H(x) =\tfrac{1}{2}\left(x+gx^3 -(m_2x^2 + m_0) \sqrt{x^2 - 4d^2} \right)
\end{align}
By imposing $G_H(x) \rightarrow 1/x$, we find $m_2=g$, $m_0 =1+2gd^2$, $d = {\sqrt{\frac{\sqrt{12 g+1}-1}{6g}}}$.

Define the Blue transform as the functional inverse of the resolvent
\begin{align}
    B(G_H(x))=x.
\end{align}
At $g= 0$, $B(y) = y + y^{-1}$.
The resolvent satisfies the quadratic equation
\begin{align}
    G_H(x)^2 - (x + g x^3) G_H(x)+(4 d^6 g^2+3 d^4 g^2 x^2+4 d^4 g+d^2 g x^2+d^2) = 0
\end{align}
We may then set $G_H(x) = y$ solve for $x$. This can be done nonperturbatively in $g$, but to leading order
\begin{align}
    B(y) = y + y^{-1}-g y \left(y^2+2\right) + O(g^2)
\end{align}

Under free compression, the Blue transform is~\cite{NicaSpeicher:2006}
\begin{align}
    B_P(x) = B(\alpha x) + \frac{\alpha -1}{\alpha x}
\end{align}
so
\begin{align}
\begin{aligned}
    &G_{PHP}(x) =\frac{x-\sqrt{x^2-4 \alpha }}{2 \alpha }
    \\&+\frac{g \left(-2 (\alpha -2) \alpha -x^4+(4 \alpha -2) x^2-2 (\alpha -1) x
   \sqrt{x^2-4 \alpha }+x^3 \sqrt{x^2-4 \alpha }\right)}{2 \alpha  \sqrt{x^2-4 \alpha }}+O\left(g^2\right)
   \end{aligned}
\end{align}
The edges of the spectrum are at $B_P(x)$ for $B_P'(x) = 0$ , so to leading order
\begin{align}
    a_P = -\sqrt{\alpha } (2-(\alpha +2) g), \quad b_P = \sqrt{\alpha } (2-(\alpha +2) g)
\end{align}

We can work perturbatively in $\epsilon = 1-\alpha$. The subordination map~\cite{Voiculescu1986,Biane1998,NicaSpeicher:2006} is
\begin{align}
    w(y)= y + \frac{\epsilon}{G(y)}+\dots
\end{align}
Then,
\begin{align}
    w'(y)  = 1-\epsilon G'(y)G(y)^{-2}+ \dots
\end{align}
and so to leading order
\begin{align}
    W_\alpha(x,y) = W_1(x,y)+ \epsilon\left(G(y)^{-1}\partial_y W_1(x,y)- G'(y)G(y)^{-2}W_1(x,y)\right) + \dots
\end{align}

We have to leading order in $1-\alpha$ and $g$
\begin{align}
    W_\alpha(x,y) = \frac{\zeta ^2 z^2 (\alpha +3 (\alpha -2) g+(\alpha +2) \zeta  g z+\zeta 
   z-2)}{\left(\zeta ^2-1\right) \left(z^2-1\right) (\zeta  z-1)^3}
\end{align}
where $x(z)=d(z+1/z)$, $y(\zeta)=d_P(\zeta+1/\zeta)$, and $d_P=b_P/2$.
And then
\begin{align}
    \begin{aligned}\la \Tr \left[T_m(\tfrac{H}{2d})\right] &\Tr\left[T_n(\tfrac{PHP}{2d_P})\right]\ra_c
    \\&= \frac{1}{(2\pi \i)^2}\oint\oint T_m(\tfrac{x(z)}{2d}) T_n(\tfrac{y(\zeta)}{2d_P})W_\alpha(x(z),y(\zeta))\d x(z)\d y(\zeta)
    \\
    &= \left(\frac{m}{4}+\frac{(\alpha -1)(1+g) m^2}{8} \right)\delta_{nm}.
    \end{aligned}
\end{align}
We can cross-check this formula by explicitly computing the first few moments 
$C_{m,n}= \Big\la \Tr H^m\;\Tr(PHP)^n\Big\ra_c$, using standard perturbation theory.
Using Mathematica, we explicitly checked for $m, n \leq 6$ that these are consistent with the general formula at $O((1-\alpha) g)$.

\section{Background and Proofs}
\label{sec:backgroundproofs}

\subsection{Proof of \eqref{eq:s0}}
\label{sec:proof}
The goal of this section is to prove \eqref{eq:s0}. To do so, we use tools from free probability theory~\cite{mingo2017free}. Define the normalized trace $\tr(\cdot) = L^{-1}\Tr(\cdot)$, which has a good large $L$ limit for the operators we consider. Namely, we consider the algebra of operators in the large-$L$ limit generated by $H$ and $P$, denoted $\mathcal{M}$ and the subalgebra generated by $H$ alone, denoted $\mathcal{B}$. We have $\tr(P) = \alpha$. 

There exists a unique trace preserving conditional expectation $\mathcal{E}$ from $\mathcal{M}$ to $\mathcal{B}$ i.e.~
\begin{align}
    \tr(\mathcal{E}(m)) = \tr(m),\quad \mathcal{E}(b_1 m b_2) = b_1 \mathcal{E}(m) b_2,\quad  m \in \mathcal{M},\quad b_1,b_2 \in \mathcal{B}.
\end{align}
{This operation was described informally in section 4 and can be defined by averaging over $P\to UPU^{-1}.$}
Any element $a\in\mathcal{M}$ can be decomposed as\footnote{Such a decomposition is unique. Suppose that we had another decomposition, different by $\tilde{b} + \sum \tilde{b}_1\dots = 0$. Applying the conditional expectation, we would find that $\tilde{b} = 0$ and thus also $\sum \tilde{b}_1\dots = 0$}
\begin{align}
    a = b + \sum_{n=2}^{\infty} b_1(P-\alpha \textbf{1})b_2\dots b_{n-1}(P-\alpha {\textbf{1}})b_{n} , \quad b_i \in \mathcal{B} \quad \tr(b_i) = 0,  \forall i \in \{2,\dots n-1\}.
\end{align}
The trace preserving conditional expectation is then $\mathcal{E}(a) = b$. The trace preserving property follows from the linearity of the trace and the fact that alternating centered elements of freely independent algebras have zero trace. We did not need to take $b_1$ or $b_n$ to be centered because, they can be decomposed into a centered piece and a piece proportional to identity so that each term in the summand becomes a sum of four terms, in each of which every element is centered. The property $\mathcal{E}(b_1 m b_2) = b_1 \mathcal{E}(m) b_2$ of the conditional expectation defined above is manifest from the decomposition.

Letting $b = (x-H)^{-1}$ and $m = P(y-PHP)^{-1}P$, then
\begin{align}
    \tr( (x-H)^{-1} P(y-PHP)^{-1}P)=  \tr( (x-H)^{-1} \mathcal{E}(P(y-PHP)^{-1}P)).
\end{align}
Corollary 4.5 of~\cite{curran2008analytic} states that there exists an analytic function $F$ such that
\begin{align}
    \alpha^{-1}\mathcal{E}((\alpha^{-1}PHP -\beta P)^{-1}) = (H-F(\beta))^{-1}.
    \label{eq:2.87}
\end{align}
The derivation is involved, extending Voiculescu's work on subordination~\cite{voiculescu1993analogues}.

We can then derive
\begin{align}
    \mathcal{E}(P(y-PHP)^{-1}P) = (\omega(y)-H)^{-1},\quad \omega(y) \equiv F(\alpha^{-1}y)
\end{align}
and so
\begin{align}
     \tr( (x-H)^{-1} P(y-PHP)^{-1}P) = \tr((x-H)^{-1}(\omega(y)-H)^{-1}).
\end{align}
Using the identity
\begin{align}
    (x-H)^{-1}-(\omega(y)-H)^{-1} = (x-H)^{-1}(\omega(y) - x)(\omega(y)-H)^{-1},
\end{align}
we find
\begin{align}
    \tr( (x-H)^{-1} P(y-PHP)^{-1}P) = \frac{G_H(x)-G(\omega(y))}{\omega(y)-x}
    \label{eq:2.91}
\end{align}
where the analytic function $\omega$ has yet to be determined.

We now define the $\mathcal{R}$-transform, not to be confused with the resolvent. The $\mathcal{R}$-transform for an operator $H$ is defined as
\begin{align}
    \mathcal{R}_H(z)= G^{(-1)}_{H}(z) -\frac{1}{z},
\end{align}
where $G^{(-1)}$ is the functional inverse of the Cauchy transform. Taking $z\rightarrow G_H(z)$ in the above formula, we see that
the $\mathcal{R}$-transform satisfies the following implicit equation
\begin{align}
\label{eq:2.42new}
    \frac{1}{G_H(z)}+\mathcal{R}_H(G_H(z)) = z.
\end{align}
What is nice about the $\mathcal{R}$-transform is that it transforms simply under free compression $H \rightarrow PHP$~\cite{NicaSpeicher:2006}
\begin{align}
    \alpha^{-1}\mathcal{R}_H(z) = \mathcal{R}_{\alpha^{-1}{PHP}}(z),
    \label{eq:2.42}
\end{align}
which is derived in section~\ref{sec:proof2.42}.
We have
\begin{align}
    \frac{1}{G_{\alpha^{-1}{PHP}}(z)}+\mathcal{R}_{\alpha^{-1}{PHP}}(G_{\alpha^{-1}{PHP}}(z) ) = z.
\end{align}
and so
\begin{align}
    \frac{1}{G_{\alpha^{-1}{PHP}}(z)}+\alpha^{-1}\mathcal{R}_{H}(G_{\alpha^{-1}{PHP}}(z) ) = z.
    \label{eq:2.95}
\end{align}
Defining
\begin{align}
    \omega_\alpha(z) \equiv \frac{1}{G_{\alpha^{-1}{PHP}}(z)}+\mathcal{R}_{H}(G_{\alpha^{-1}{PHP}}(z) )
\end{align}
and using the definition of the $\mathcal{R}$-transform for $H$ itself
\begin{align}
    \frac{1}{G_{H}(\omega_\alpha(z))}+\mathcal{R}_{H}(G_{H} (\omega_\alpha(z))) = \omega_\alpha(z),
\end{align}
we see that
\begin{align}
    G_H(\omega_{\alpha}(z)) = G_{\alpha^{-1}{PHP}}(z)
    \label{eq:2.98}
\end{align}
From \eqref{eq:2.95}, we have that
\begin{align}
    z&=\frac{1}{G_{\alpha^{-1}{PHP}}(z)}+\mathcal{R}_{H}(G_{\alpha^{-1}{PHP}}(z) )+(\alpha^{-1}-1)\mathcal{R}_{H}(G_{\alpha^{-1}{PHP}}(z) ) 
    \\
    &= \omega_\alpha (z) +(\alpha^{-1}-1)\mathcal{R}_{H}(G_{\alpha^{-1}{PHP}}(z) ) 
    \\
    &= \omega_\alpha (z) +(\alpha^{-1}-1)( \omega_\alpha (z)- \frac{1}{G_{\alpha^{-1}{PHP}}(z)}).
\end{align}
Taking $z = \alpha^{-1}y$,
we have
\begin{align}
    \omega_\alpha(\alpha^{-1}y)= y + \frac{1-\alpha}{G_H(\omega_{\alpha}(\alpha^{-1}y))}
\end{align}
Relating the trace of \eqref{eq:2.87} and the equality of \eqref{eq:2.98}, we identify $F(z)$ with $\omega_\alpha(z)$ and so 
\begin{align}
    \omega(y)= y + \frac{1-\alpha}{G_H(\omega(y))}
\end{align}
Together with \eqref{eq:2.91}, this completes the proof of \eqref{eq:s0}.

\subsection{Proof of \eqref{eq:2.42}}
\label{sec:proof2.42}

In this subsection, we prove the transformation property of the $\mathcal{R}$-transform under free compression. First, we recall that algebras $\mathcal{X}$ and $\mathcal{Y}$ are freely independent with respect to the state $\tr$ if all alternating centered moments are zero
\begin{align}
    \tr(x_1y_1\dots x_n y_n) = 0, \quad \forall x_i \in \mathcal{X}, \quad y_i \in \mathcal{Y} \text{ s.t. } \tr(x_i) = \tr(y_i) = 0.
\end{align}

We now define the free cumulants, $\kappa_n$. These may be recursively derived using the moment-cumulant relation
\begin{align}
    \tr(a_1 \dots a_n) \equiv \sum_{\pi \in NC_n} \kappa_{\pi}(a_1,\dots, a_n), \quad a_i \in \mathcal{A}
\end{align}
This definition requires some unpacking. First, $NC_n$ is the set of noncrossing partitions of $n$ elements. $\kappa_{\pi}$ is the product of free cumulants, one for each block of the partition $\pi$. We also note that this formula does not assume any algebraic structure between the $a_i$'s such as freeness. To orient ourselves, let us work out the first few free cumulants. For $n = 1$, there is a single non-crossing partition, the identity element and we find
\begin{align}
    \tr(a) = \kappa_1(a).
\end{align}
For $n = 2$, we have both $(1)(2)$ and $(1,2)$, so
\begin{align}
    \tr(a_1a_2) = \kappa_1(a_1)\kappa_1(a_2) + \kappa_2(a_1,a_2) = \tr(a_1)\tr(a_2) + \kappa_2(a_1,a_2)
\end{align}
thus
\begin{align}
    \kappa_2(a_1,a_2) = \tr(a_1a_2) -\tr(a_1)\tr(a_2).
\end{align}
We can then continue recursively. This becomes different than classical cumulants at $n = 4$ because that is where the set of noncrossing partitions is distinct from the set of all partitions.

We now prove that the $\mathcal{R}$-transform is the generating function of free cumulants
\begin{align}
    \mathcal{R}_a(z) = \sum_{n \geq 1}\kappa_n(a,\dots,a)z^{n-1}
\end{align}
Define $m_n = \tr(a^n)$, then the moment cumulant relation asserts that
\begin{align}
    m_n = \sum_{\pi \in NC_n}\kappa_\pi(a,\dots,a).
\end{align}
Within each $\pi$, there is a block that contains ``$1$'' that we denote $V$ and this block can be of any length between $1$ and $n$. Because $\pi$ is noncrossing, the other blocks in $\pi$ must be in the gaps (of lengths $i_1, i_2,\dots i_k$) between the elements of $V$. Thus, for each gap of length $i_j$, there is a sum of noncrossing partitions of $i_j$ elements with the associated free cumulants. By the moment cumulant relation for $m_{i_j}$, this simply gives a factor of $m_{i_j}$, so summing over all possibilities
\begin{align}
    m_n = \sum_{k = 1}^n \kappa_k(a,\dots,a) \sum_{i_1+\dots+i_k = n-k}m_{i_1}\dots m_{i_k}.
    \label{eq:2.60}
\end{align}
Defining
\begin{align}
    M_a(z) = \sum_{n = 1}^{\infty} m_nz^n,
\end{align}
we have 
\begin{align}
    1+M_a(z) = \sum_{n = 0}^{\infty} m_nz^n
\end{align}
and
\begin{align}
    (1+M_a(z))^k= \sum_{i_j= 0}^{\infty} m_{i_1}\dots m_{i_k }z^{i_1+\dots+i_k}
\end{align}
and so 
\begin{align}
    \sum_{i_1+\dots+i_k = n-k}m_{i_1}\dots m_{i_k} = [z^{n-k}](1+M_a(z))^k
\end{align}
where $[z^{n-k}]$ means the $n-k$ coefficient.
Multiplying \eqref{eq:2.60} by $z^n$ and summing
\begin{align}
\begin{aligned}
    M_a(z) = \sum_{n = 1}^{\infty}\sum_{k = 1}^n \kappa_k (a,\dots,a) z^n [z^{n-k}](1+M_a(z))^k
    \end{aligned}
\end{align}
Taking $r = n -k$
\begin{align}
\begin{aligned}
    M_a(z) =  \sum_{k = 1}^{\infty} \kappa_k(a,\dots,a) \sum_{r=0}^{\infty} z^{r+k} [z^{r}](1+M_a(z))^k =  \sum_{k = 1}^{\infty}  \kappa_k(a,\dots,a) (1+M_a(z))^k z^k 
    \\= C_a(z(1+M_a(z))),
    \label{eq:2.66}
    \end{aligned}
\end{align}
where 
\begin{align}
    C_a(z) \equiv \sum_{n = 1}^{\infty}\kappa_n(a,\dots,a)z^n 
\end{align}
$M_a(z)$ is related to the Cauchy transform as
\begin{align}
    M_a(1/z) = \sum_{n = 1}^{\infty} m_n z^{-n} = z\sum_{n = 0}^{\infty} m_n z^{-n-1} - 1 = z G_a(z) -1
\end{align}
and from \eqref{eq:2.66}
\begin{align}
    M_a(1/z) = C_a(z^{-1}(1+M(z^{-1}))) = C_a(G_a(z))
\end{align}
so
\begin{align}
    z G_a(z) -1 = C_a(G_a(z))
\end{align}
Defining 
\begin{align}
    \tilde{\mathcal{R}}_a(z) = z^{-1}C_a(z) = \sum_{n = 1}^{\infty}\kappa_n(a,\dots,a) z^{n-1}
\end{align}
we have
\begin{align}
     z  -\frac{1}{G_a(z)} = \tilde{\mathcal{R}}_a(G_a(z))
\end{align}
which is the implicit equation satisfied by the $\mathcal{R}$-transform \eqref{eq:2.42new}. Thus 
\begin{align}
    \mathcal{R}_a(z) =  \sum_{n = 1}^{\infty}\kappa_n(a,\dots,a) z^{n-1}
\end{align}
as claimed.

A key fact about free cumulants is that if $\mathcal{X}$ and $\mathcal{Y}$ are freely independent, then all mixed cumulants vanish. A mixed cumulant is one with arguments coming from both $\mathcal{X}$ and $\mathcal{Y}$. This can be proved by induction. For $n = 1$, there is no way to have a mixed cumulant and for $n = 2$, it is straightforward to check that the definition of freeness shows that the mixed cumulant vanishes. Now assume that for all orders below $n$, the mixed cumulants vanish. We begin by focusing on mixed cumulants of alternating centered variables, so by the definition of freeness, the mixed moment is zero, and so by the moment cumulant formula
\begin{align}
    0 = \sum_{\pi\in NC_n} \kappa_\pi(a_1,\dots,a_n).
\end{align}
Due to the alternating property, every $\pi$ either contains a $\kappa_1$, which is zero by centeredness or contains a mixed cumulant of order smaller than $n$, which is zero by the induction assumption, or is $\kappa_n$. Thus $\kappa_n = 0$. Next, we remove the centeredness restriction. By induction, it is straightforward to prove that any $\kappa_n$ with $n\geq 2$ and at least one argument equal to the identity operator is zero. Then, by the multilinearity of $\kappa_n$, we find that $\kappa_n(a_1,\dots, a_n) = \kappa_n(a_1- \tr(a_1) \textbf{1},\dots, a_n-\tr(a_n) \textbf{1})=0$. Finally, we need to remove the alternating restriction. To do so, we use the following identity of free cumulants, which is Theorem 11.12 of~\cite{NicaSpeicher:2006}
\begin{align}
    \kappa_{n-1}(a_1,\dots,a_k a_{k+1},\dots,a_n) = \sum_{\pi \in NC_n, \pi \vee \{k,k+1\} = \{1,\dots,n\}} \kappa_\pi(a_1,\dots, a_n)
\end{align}
where the sum is over partitions where, upon merging the blocks with $k$ and $k+1$, the partition is the full set. We have assumed that the $a_i$'s are not alternating, so we merge $a_k$ and $a_{k+1}$ that are from the same algebra. By the induction assumption, the left hand side of the equation is zero. In the sum, every $\pi$ satisfying the property $\pi \vee \{k,k+1\} = \{1,\dots,n\}$ has a block with mixed cumulants, and so by the induction assumption, these are all zero, except for when $\pi = \{1,\dots, n\}$ which has yet to be determined. However, all other terms are zero, so we conclude that $\kappa_n(a_1,\dots,a_n) = 0$, completing the proof.

From the vanishing of mixed cumulants
\begin{align}
    \tr(b_1Pb_2P\dots b_n P) &= \sum_{\pi \in NC_{2n}}\kappa_\pi(b_1,P,\dots,b_n,P) 
    \\
    &= \sum_{\substack{\pi_b \in NC_{(1,3,\dots,2n-1)}
    \\
    \pi_P \in NC_{(2,4,\dots,2n)}
    \\
    \pi_b\cup \pi_P\in NC_{2n}}}\kappa_{\pi_b}(b_1,\dots,b_n)\kappa_{\pi_P}(P,\dots,P)
\end{align}
There is a largest element $\pi_P \in NC_{(2,4,\dots,2n)}$ and $\pi_b\cup \pi_P\in NC_{2n}$ that we call $K(\pi_b)$, the so-called Kreweras complement, thus
\begin{align}
     \tr(b_1Pb_2P\dots b_n P) &= \sum_{\pi_b \in NC_n}\kappa_{\pi_b}(b_1,\dots,b_n) \sum_{\pi_P \leq K(\pi_b)}\kappa_{\pi_P}(P,\dots,P) \\
     &= \sum_{\pi_b \in NC_n}\kappa_{\pi_b}(b_1,\dots,b_n) \tr(P)^{|K(\pi_b)|}
     \label{eq:2.63}
\end{align}
where $|K(\pi_b)|$ is the number of blocks of $K(\pi_b)$. In this second equality, we used the moment cumulant relation $|K(\pi_b)|$ times and the fact that $P$ is a projector.

Using \eqref{eq:2.63},
\begin{align}
    \tr(Pb_1Pb_2P\dots b_n P) &= \sum_{\pi \in NC_{n+1}}\kappa_{\pi}(\textbf{1},b_1,\dots,b_n) \alpha^{|K(\pi)|}
    \\
    &= \sum_{\pi \in NC_{n}}\kappa_{\pi}(b_1,\dots,b_n) \alpha^{|K(\pi)|}
\end{align}
From the moment-cumulant relation and the vanishing of mixed cumulants (noting $\textbf{1}$ is freely independent from everything), we see that $\kappa_{\pi}(\textbf{1},b_1,\dots,b_n) = 0$ when $\pi$ is not of the form $(0)\cup \tilde{\pi}$ and otherwise $\kappa_{\pi}(\textbf{1},b_1,\dots,b_n) =\kappa_{\tilde{\pi}}(b_1,\dots,b_n)$. Moreover, for all $\pi \in NC_{n+1}$, $|\pi| + |K(\pi)| = n+2$ and $|\pi| = |\tilde{\pi}| +1$, thus
\begin{align}
    \tr(Pb_1Pb_2P\dots b_n P) &=  \sum_{\pi \in NC_{n}}\kappa_{\pi}(b_1,\dots,b_n) \alpha^{n+1-|\pi|} = \sum_{\pi \in NC_{n}}\kappa_{\pi}(\alpha b_1,\dots,\alpha b_n) \alpha^{1-|\pi|} 
\end{align}
From the moment cumulant relation in the compressed theory
\begin{align}
    \alpha^{-1}\tr(Pb_1Pb_2P\dots b_n P) = \sum_{\pi\in NC_n}\kappa_\pi^{P\mathcal{B}P}(Pb_1P,\dots,Pb_nP).
\end{align}
The superscript on the cumulant denotes that we are using the $\alpha^{-1}$ normalization on the left hand side, so that $\kappa_1^{P\mathcal{B}P}(P) =1$.
Comparing these formulas, we have
\begin{align}
    \kappa_\pi^{P\mathcal{B}P}(Pb_1P,\dots,Pb_nP) = \kappa_{\pi}(\alpha b_1,\dots,\alpha b_n) \alpha^{-|\pi|} 
\end{align}
and thus
\begin{align}
    \kappa_n^{P\mathcal{B}P}(Pb_1P,\dots,Pb_nP) = \alpha^{-1}\kappa_n(\alpha b_1, \dots, \alpha b_n), \quad b_i \in \mathcal{B}.
\end{align}
Using its representation as the generating function of free cumulants, the $\mathcal{R}$-transform for $\alpha^{-1}PHP$ is thus
\begin{align}
\begin{aligned}
    \mathcal{R}_{\alpha^{-1}PHP }(z)  = \sum_{n = 0}^{\infty}\kappa_{n+1}^{P\mathcal{B}P }(\alpha^{-1}PHP, \dots, \alpha^{-1}PHP)z^n =\sum_{n = 0}^{\infty}\alpha^{-1}\kappa_{n+1}(H,\dots,H)z^n  
    \\
    =\alpha^{-1} \mathcal{R}_{H}(z),
    \end{aligned}
\end{align}
completing the derivation of \eqref{eq:2.42}.

\bibliographystyle{JHEP}
\bibliography{main}
\end{document}